\documentclass[10pt,journal,final]{IEEEtran}
\usepackage{amsmath, amssymb, amsthm}
\usepackage{graphicx, graphics, epsfig, color}
\usepackage{subfigure}
\usepackage{tikz, pgfplots}
\usepackage{algorithm, algorithmic}
\usepackage{float}
\usepackage{multirow}
\usepackage{cuted, flushend}
\usepackage{midfloat}
\usepackage{bm}
\usepackage{fancyhdr}
\usepackage{enumerate}
\usepackage[numbers,square,sort&compress]{natbib}
\usepackage{epstopdf}
\usepackage{mathtools}

\newtheorem{theorem}{Theorem}
\newtheorem{cor}{Corollary}

\newtheorem{prop}{Proposition}

\DeclareMathOperator*{\argmax}{argmax}

\pgfplotsset{compat=1.4}
\vspace{4ex}

\usepackage[symbol]{footmisc}
\graphicspath{{figs//}}

\begin{document}
\title{{Hypothesis Test Procedures for Detecting Leakage Signals in Water Pipeline Channels}}
\author{Liusha Yang, Matthew R. McKay, Xun Wang}
\date{}
\maketitle
\fussy

\begin{abstract}
We design statistical hypothesis tests for performing leak detection in water pipeline channels. By applying an appropriate model for signal propagation, we show that the detection problem becomes one of distinguishing signal from noise, with the noise being described by a multivariate Gaussian distribution with unknown covariance matrix. We first design a test procedure based on the generalized likelihood ratio test, which we show through simulations to offer appreciable leak detection performance gain over conventional approaches designed in an analogous context (for radar detection).  Our proposed method requires estimation of the noise covariance matrix, which can become inaccurate under high-dimensional settings, and when the measurement data is scarce.  To deal with this, we present a second leak detection method, which employs a regularized covariance matrix estimate. The regularization parameter is optimized for the leak detection application by applying results from large dimensional random matrix theory.  This second proposed approach is shown to yield improved performance in leak detection compared with the first approach, at the expense of requiring higher computational complexity.
\end{abstract}

\begin{IEEEkeywords}
Leak detection, hypothesis test, random matrix theory.
\end{IEEEkeywords}

\section{Introduction}
Leakage in water supply systems causes wastage of water and energy resources, and poses public health risk due to water pollution. Leaks may occur, for example, due to aging pipelines, corrosion, and excessive steady and/or unsteady pressures in the system \cite{ghazali2012comparative}. Thus, an effective leakage detection method is essential.

Most related research in this area has focused on the problem of leak  estimation, for which the objective is usually to estimate the location of the leak, assuming that a leak actually exists in the pipeline. For this purpose, various transient-based leak location estimation methods have been developed
(e.g., \cite{ghazali2012comparative,vitkovsky2000leak,wang2018pipeline,wang2018identification}).
In this work we address the related (but different) problem of leak detection, by developing suitable statistical hypothesis testing procedures. Despite being a natural detection approach, to our knowledge, hypothesis tests have yet to be developed for leak detection in water pipeline systems.

Generally speaking, we develop data-driven approaches to decide between the presence or absence of a leak in the pipeline, and for the former case, return estimates of the leak parameters. The measured data corresponds to primary and secondary measurements of head differences at different frequencies, taken from multiple sensors deployed at different locations along the water pipeline. Our tests are developed based on a linearized transient wave model in the frequency domain, as proposed in \cite{wang2018pipeline,wang2018identification}, which has been supported by experimental data \cite{wang2018experiment}. By applying hypothesis testing theory to this model, we find that, from a technical point of view, the problem boils down to a binary classification problem that discriminates between a ``null hypothesis'', corresponding to zero-mean complex Gaussian noise with non-trivial correlation, and an ``alternative hypothesis'', corresponding to a structured (deterministic) signal embedded within the Gaussian noise.  For the latter hypothesis, the deterministic signal is a function of the leak parameters, including size and location.

Since the signal and noise model parameters (i.e., noise covariance, leak location and size) are all unknown, we develop test procedures based on the generalized likelihood ratio test (GLRT) \cite{kelly1986adaptive}, which constructs a likelihood ratio based on the two hypotheses, and replaces the unknown parameters in the likelihood functions by appropriate estimates. We first consider a traditional strategy of which replaces the unknown parameters by their maximum likelihood estimates (MLE), and develop a suitable test statistic.  This statistic exploits the known structure of the leak signals (under the alternative hypothesis), and is proven to have the desirable property of being a constant false alarm rate (CFAR) statistic; meaning that a detection threshold can be specified which achieves a fixed false alarm probability, regardless of the model parameters.  Through simulations, we demonstrate the good performance of the proposed method in detecting leaks, and show enhancement over methods that have been developed for related models in the context of radar detection.  This approach is particularly suited to ``data rich'' scenarios, where the MLEs provide accurate parameter estimates.

One limitation of the proposed approach is that for high dimensional settings when the number of frequency domain measurements and/or the number of sensors is large, the number of parameters to estimate is also large.  This is particularly the case for the noise covariance matrix, and it is well known that under high dimensional settings that the MLE -- corresponding to the conventional sample covariance matrix (SCM) estimate -- is particularly inaccurate.  This, in turn, can degrade the performance of the proposed leak detection algorithm. To deal with this potential problem, we propose a second detection algorithm that seeks to design a robust covariance estimation solution which is suitably optimized for the task of leak detection, under high dimensional settings.  The approach is to replace the SCM with a regularized version (termed RSCM) in the GLRT statistic, and to optimize the regularization parameter to maximize the leak detection accuracy subject to a prescribed false alarm criteria. The RSCM is a simple but effective covariance matrix estimator to deal with problems of sample deficiency and high dimensionality by pulling the spread sample eigenvalues toward their grand mean \cite{ledoit2004well}.
It is used in many fields, including mathematical finance and adaptive array processing \cite{rubio2012performance,abramovich1981controlled,abramovich2007modified,mestre2006finite,carlson1988covariance}. Extensions have also been proposed which replace the SCM with a robust covariance matrix estimator (such as Tyler's estimator) to provide resilience against outliers \cite{couillet2014large,yang2015robust,auguin2018large}.  The main challenge is generally to develop data-driven methods to optimize the regularization parameter, which is typically application dependent. In a similar spirit to previous work (e.g., \cite{ledoit2004well,mestre2006finite,rubio2012performance,ma2003efficient,kammoun2018optimal,couillet2016second}), our solution draws from recent results in the area of large dimensional random matrix theory.  Most specifically, it leverages technical results from \cite{couillet2016second,kammoun2018optimal}, which considered a related detection problem, but which considered a different model to the one in this paper.

The basic idea of the approach is to first characterize the asymptotic behavior of the false alarm and detection probabilities under certain double-limit asymptotics, which we define, and subsequently to provide consistent estimators of these probabilities which are completely data-driven. Based on this, we can then optimize the regularization parameter in an online fashion, which maximizes the (estimated) detection probability while maintaining a prescribed (estimated) false alarm probability. The performance of this second proposed leak detection algorithm is demonstrated through simulations, and shown to outperform the first proposed algorithm, particularly under high-dimensional model settings, at the expense of increased complexity.

\section{System model}\label{sec:model}
\begin{figure}[!htb]
\begin{center}
\includegraphics[width=1\linewidth]{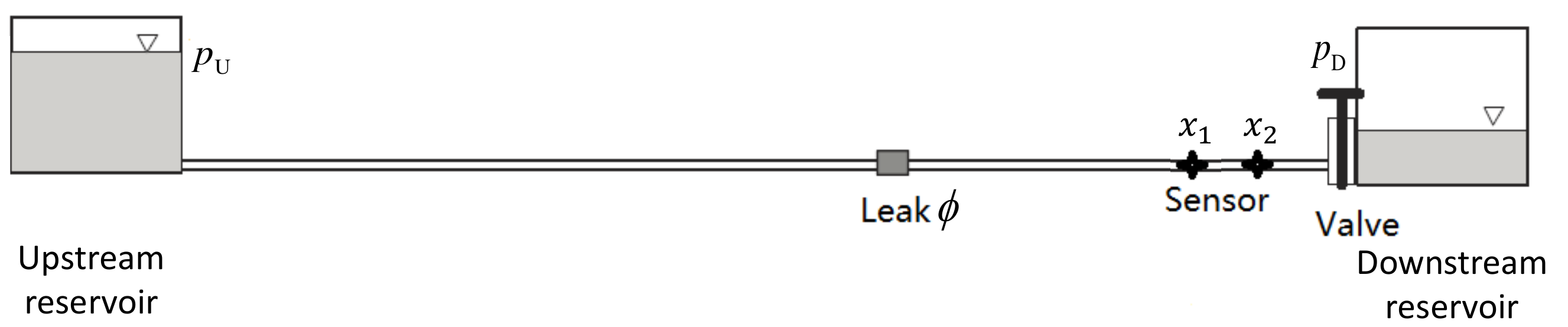}
\caption{Pipeline configuration. Under hypothesis $H_0$, there is no leak in the water pipe, while under hypothesis $H_1$, a leak of size $s$ is present at location $\phi$. }
\label{fig:PipelineSystemConfigH1}
\end{center}
\vspace{-0.3cm}
\end{figure}

As shown in Fig.\,\ref{fig:PipelineSystemConfigH1}, we consider a reservoir-pipe-valve system where the pipe of length $l$ meters is bounded by $p_{\rm U}=0$ and $p_{\rm D}=l$. A total of $M$ pressure sensors deployed near the downstream node are used to collect pressure head oscillations\footnote{The pressure head (in meters) relates the pressure of a fluid to the height of a column of that fluid having an equivalent static pressure at its base. The head is defined as $h=p/(\rho g)$ where $p$ is the pressure (in Pascals), $g$ denotes gravitational acceleration, and $\rho$ is the density of the fluid. For example, 50 m of head in a pipe implies that if that pipe bursts, the height of the resulting water jet would be 50 m.} for leak identification.  The locations of the $M$ sensors are $p_{\rm U}<x_1<x_2<\ldots<x_M<p_{\rm D}$. We denote the leak size and the leak location as $s$ and $\phi$.

By rapidly closing and/or opening the valve at the downstream of the pipe, the sensors measure the pressure head oscillations at different frequencies, which are affected by a leak in the pipe. Let $h_m(w_j)$ denote the head oscillation at frequency $w_j$ and location $x_m$, and $h^o_m(w_j)$ the computed head oscillation with no leak, where $j=1,\ldots,J$ and $m=1,\ldots,M$. We define the head difference at frequency $w_j$ observed by the sensor at $x_m$ as $z_m(w_j)=h_m(w_j)-h^o_m(w_j)$. If the pipe is intact (with no leak), $z_m(w_j)=h_m(w_j)-h^o_m(w_j)=n_m(w_j)$, where $n_m(w_j)$ is the measurement noise, which can be measurement error or environment noise induced by turbulence, traffic, construction, etc. Otherwise, $z_m(w_j)=sg_m(\phi, w_j)+n_m(w_j)$, in which $sg_m(\phi, w_j)$ is the leak component, which depends on the leak size $s$ and the leak location $\phi$. The detailed formulas of $h^o_m(w_j)$ and $g_m(\phi, w_j)$ are provided in the Appendix \ref{appx:model}. Assembling $z_m(w_j)$ into a vector ${\bf z}_0\in\mathbb{C}^N$ of length $N=J\times M$, we have
 \begin{align}\nonumber
 {\bf z}_0={\rm vec}[z_m(w_j), j=1,\ldots,J, m=1,\ldots,M].
 \end{align}
 We denote the hypothesis of whether there exists a leak or not by $H_1$ and $H_0$, respectively.
 Then the problem of detecting a leak in a noise-contaminated water pipe can be posed in terms of the following binary hypothesis test:
 \begin{align}\label{eq:H1H0}
\left\{
\begin{array}{l}
H_0: {\bf z}_0={\bf n}_0,  \\
H_1: {\bf z}_0=s{\bf g}(\phi)+{\bf n}_0
\end{array}
\right.
\end{align}
where the noise vector ${\bf n}_0={\rm vec}[n_m(w_j)]$ is assumed to be Gaussian distributed\footnote{The Gaussian noise assumption in water pipes with flow is justified by experimental investigations in laboratory pipe systems \cite{dubey2019measurement}.} with zero mean and covariance matrix ${\bf C}_N$, and
\begin{align}\nonumber
{\bf g}(\phi)={\rm vec}[g_m(\phi,w_j), j=1,\ldots,J, m=1,\ldots,M].
\end{align}

We assume that $K$ independent samples of noise-only data are available, which are referred to as secondary data:
\begin{align}\nonumber
{\bf z}_k={\bf n}_k, ~{\bf n}_k\sim CN(\textbf{0}, {\bf C}_N), ~k=1,\ldots,K.
\end{align}
These may be obtained, for example, by the steady-state pressure measurements when the pipe is newly built.


Thus, the leak detection problem can be recast as the following hypotheses:
\begin{align}\nonumber
\left\{
\begin{array}{l}
H_0: {\bf z}_0={\bf n}_0, \quad\quad\quad\quad\,\,\;{\bf z}_k={\bf n}_k, ~~k=1\ldots,K\\
H_1: {\bf z}_0=s{\bf g}(\phi)+{\bf n}_0, \quad{\bf z}_k={\bf n}_k, ~~k=1\ldots,K.
\end{array}
\right.
\end{align}
The joint probability density function (PDF) of the input data under $H_0$ is
\begin{align}\nonumber
&f_0({\bf z}_0,\ldots,{\bf z}_K|H_0)=\\ \label{eq:pdf_f0}
&\frac{1}{(\pi^N\det({\bf C}_N))^{K+1}}{\rm exp}\left[-\sum_{k=1}^K{\bf z}_k^H{\bf C}_N^{-1}{\bf z}_k\right]{\rm exp}\left[-{\bf z}_0^H{\bf C}_N^{-1}{\bf z}_0\right]
\end{align}
where $\det({\bf C}_N)$ is the matrix determinant of ${\bf C}_N$.

Similarly, the joint PDF of the input data under $H_1$ is
\begin{align} \nonumber
&f_1({\bf z}_0,\ldots,{\bf z}_K|H_1)=\\ \nonumber
&\frac{1}{(\pi^N\det({\bf C}_N))^{K+1}}{\rm exp}\left[-\sum_{k=1}^K{\bf z}_k^H{\bf C}_N^{-1}{\bf z}_k\right]\\ \label{eq:pdf_f1}
&\times{\rm exp}\!\left[-({\bf z}_0-s{\bf g}(\phi))^H{\bf C}_N^{-1}({\bf z}_0-s{\bf g}(\phi))\right].
\end{align}

The most natural approach to detect the presence of a leak is the likelihood ratio (LR) test, which computes the LR or its logarithm and compares it with a certain threshold $\alpha$ \cite{lehmann2006testing}. Specifically, the LR test is
\begin{align}\nonumber
L=\frac{f_1({\bf z}_0,\ldots,{\bf z}_K|H_1)}{f_0({\bf z}_0,\ldots,{\bf z}_K|H_0)}\mathop{\gtrless}^{H_1}_{H_0}\alpha.
\end{align}
Namely, if $L>\alpha$, we decide $H_1$, and if $L\leq\alpha$, we decide $H_0$.

 The LR test is known to maximize the detection probability $P_{\rm D}$ at a certain false alarm probability $P_{\rm FA}$. The $P_{\rm D}$ is defined as the probability that the detector correctly decides hypothesis $H_1$:
\begin{align}\nonumber
P_{\rm D}=\mathbb{P}[L>\alpha|H_1],
\end{align}
and the $P_{\rm FA}$ is defined as the probability that the detector decides hypothesis $H_1$ when the true hypothesis is $H_0$:
\begin{align} \label{eq:Pfa}
P_{\rm FA}=\mathbb{P}[L>\alpha|H_0].
\end{align}

For leak detection in a water pipeline system, we usually do not know the parameters $s$, $\phi$ and ${\bf C}_N$ in the PDFs $f_0({\bf z}_0,\ldots,{\bf z}_K|H_0)$ and $f_1({\bf z}_0,\ldots,{\bf z}_K|H_1)$. In this context, the LR test can not be employed. The GLRT, which employs the MLEs of the unknown parameters, is a suitable solution.

\section{Generalized likelihood ratio test (GLRT)}
In this section, we derive a GLRT-based leak detection approach and demonstrate its desirable CFAR property. The performance of our proposed approach is also assessed by numerical simulations.
\subsection{Derivation of GLRT}
We denote the leak component in the data model as ${\bf p}=s{\bf g}(\phi)$ and assume that $K\geq N$. By estimating $s$ and $\phi$, we get the estimate of ${\bf p}$. The considered GLRT is
\begin{align}\label{eq:test}
L=\frac{\max_{s, \phi}\max_{{\bf C}_N}f_1({\bf z}_0,\ldots,{\bf z}_K|H_1)}{\max_{{\bf C}_N}f_0({\bf z}_0,\ldots,{\bf z}_K|H_0)}\mathop{\gtrless}^{H_1}_{H_0}\alpha.
\end{align}
The MLEs of ${\bf C}_N$ under $H_0$ and $H_1$ are equal to the SCM, which are well known \cite{goodman1963statistical}.  Namely, the MLE of ${\bf C}_N$ under $H_0$ is $\frac{1}{K+1}\sum_{k=0}^K{\bf z}_k{\bf z}_k^H$
and the MLE of ${\bf C}_N$ under $H_1$ is \\
$\frac{1}{K+1}\left[({\bf z}_0-s{\bf g}(\phi))({\bf z}_0-s{\bf g}(\phi))^H+\sum_{k=1}^K{\bf z}_k{\bf z}_k^H\right]$.

Denote ${\bf S}_N=\sum_{k=1}^K{\bf z}_k{\bf z}_k^H$. Following similar derivation steps in \cite{kelly1986adaptive}, we obtain the MLEs of $s$ and $\phi$:
\begin{align} \label{eq:sLR}
\hat{s}=\frac{{\rm Re}\{{\bf g}^H(\phi){\bf S}_N^{-1}{\bf z}_0\}}{{\bf g}^H(\phi){\bf S}_N^{-1}{\bf g}(\phi)},
\end{align}
and the MLE of $\phi$ is
\begin{align}\label{eq:gR}
\hat{\phi}=\argmax_{\phi\in[p_{\rm U}, p_{\rm D}]}\frac{{\rm Re}^2\{{\bf g}^H(\phi){\bf S}_N^{-1}{\bf z}_0\}}{{\bf g}^H(\phi){\bf S}_N^{-1}{\bf g}(\phi)}.
\end{align}

The statistic in (\ref{eq:gR}) can be seen as a generalization of the leak location estimator presented in \cite{wang2018pipeline,wang2018identification}, which considered the problem of leak estimation under white Gaussian noise.  Because of the complicated structure of ${\bf g}(\phi)$, it is not easy to obtain an explicit formula for $\hat{\phi}$ from (\ref{eq:gR}), unlike for the other parameters. Thus, we obtain the MLE of $\phi$ through a grid search in the range $[p_{\rm U}, p_{\rm D}]$ that minimizes $\lambda$.

By plugging the MLEs of $s$, $\phi$ and ${\bf C}_N$, the test (\ref{eq:test}) becomes
\begin{align}\label{eq:test_complex}
\frac{1+{\bf z}_0^H{\bf S}_N^{-1}{\bf z}_0}{1+{\bf z}_0^H{\bf S}_N^{-1}{\bf z}_0-\frac{{\rm Re}^2\{{\bf g}^H(\hat{\phi}){\bf S}_N^{-1}{\bf z}_0\}}{{\bf g}^H(\hat{\phi}){\bf S}_N^{-1}{\bf g}(\hat{\phi})}}\mathop{\gtrless}^{H_1}_{H_0}\alpha.
\end{align}
Denote $\alpha_1=1-\frac{1}{\alpha}$. The hypothesis test (\ref{eq:test_complex}) can be further simplified as
\begin{align}\label{eq:propose}
\Delta=\frac{{\rm Re}^2\{{\bf g}^H(\hat{\phi}){\bf S}_N^{-1}{\bf z}_0\}}{(1+{\bf z}_0^H{\bf S}_N^{-1}{\bf z}_0){\bf g}^H(\hat{\phi}){\bf S}_N^{-1}{\bf g}(\hat{\phi})}\mathop{\gtrless}^{H_1}_{H_0}\alpha_1.
\end{align}

Similar to the GLRT in \cite{kelly1986adaptive}, the distribution of the test statistic $\Delta$ under $H_0$ is independent of ${\bf C}_N$ and ${\bf g}(\hat{\phi})$. Hence, the cumulative distribution function (CDF) of $\Delta$ under $H_0$, denoted as $F_{\Delta}$, remains the same for any covariance matrix ${\bf C}_N$ and nonzero vector ${\bf g}(\hat{\phi})$. Consequently, although a closed-form expression of $F_{\Delta}$ is difficult to derive, it is sufficient to apply Monte-Carlo simulations to obtain the empirical CDF $F_{\Delta}$ based on simulated data by setting ${\bf g}(\hat{\phi})=[1,0,\ldots,0]^T$, ${\bf C}_N={\bf I}_N$ and generating ${\bf z}_i$, $i=0,\ldots, K$ as standard normal distributed random vectors. The threshold $\alpha_1$ for a desired $P_{\rm FA}$ can then be determined by computing $\alpha_1=F_\Delta^{-1}(1-P_{\rm FA})$.

Although the GLRT in (\ref{eq:test}) is similar to that in \cite{kelly1986adaptive}, we should point out the main differences between the two GLRTs. Firstly, in (\ref{eq:test}), the leak size $s$ is confined to be a real number but in \cite{kelly1986adaptive}, $s$ is complex, which leads to a different MLE expression of $s$ as in (\ref{eq:sLR}). Additionally, while in \cite{kelly1986adaptive}, the signal vector ${\bf g}$ is known, in our case, ${\bf g}$ is parameterized by unknown leak location $\phi$, which is estimated in (\ref{eq:gR}).

The detection procedure is summarized in {\bf Algorithm\;\ref{al:alg1}}. As the detection test (\ref{eq:propose}) uses the SCM as the estimate of ${\bf C}_N$, we refer to this leak detection (LD) scheme as LD-SCM.
\begin{algorithm}[h]
{\small{
\caption{LD-SCM}
\label{al:alg1}
\begin{enumerate}
\smallskip
\item Determine the threshold $\alpha_1$ corresponding to the prescribed $P_{\rm FA}$ and the empirical CDF $F_{\Delta}$:
\begin{align}\nonumber
\alpha_1=F_\Delta^{-1}(1-P_{\rm FA}).
\end{align}
\item Find the optimal estimate of $\phi$ and thus ${\bf g}(\hat{\phi})$ by numerically solving:
\begin{align}\label{eq:phi_est}
\hat{\phi}=\argmax_{\phi\in[p_{\rm U}, p_{\rm D}]}\frac{{\rm Re}^2\{{\bf g}^H(\phi){\bf S}_N^{-1}{\bf z}_0\}}{{\bf g}^H(\phi){\bf S}_N^{-1}{\bf g}(\phi)}.
\end{align}
\item Compute the test statistic:
\begin{align}\nonumber
\Delta=\frac{{\rm Re}^2\{{\bf g}^H(\hat{\phi}){\bf S}_N^{-1}{\bf z}_0\}}{(1+{\bf z}_0^H{\bf S}_N^{-1}{\bf z}_0){\bf g}^H(\hat{\phi}){\bf S}_N^{-1}{\bf g}(\hat{\phi})}.
\end{align}
\item Accept $H_0$ (``no leak"), if $\Delta\leq\alpha_1$; otherwise accept $H_1$ (``leak present").
\item If $H_1$ accepted, set the estimates of $\phi$ from (\ref{eq:phi_est}) and $s$:
\begin{align}\nonumber
\hat{s}=\frac{{\rm Re}\{{\bf g}^H(\hat{\phi}){\bf S}_N^{-1}{\bf z}_0\}}{{\bf g}^H(\hat{\phi}){\bf S}_N^{-1}{\bf g}(\hat{\phi})}.
\end{align}
\end{enumerate}
}}
\end{algorithm}

\subsection{Performance evaluation and comparison}\label{sec:sim_3methods}
Here we demonstrate the performance of our proposed LD-SCM scheme, and compare it against alternative detection methods. The system configuration is shown in Fig.\;\ref{fig:PipelineSystemConfigH1}. A water pipe in a horizontal plane with length $l=2000$ m and diameter $D=0.5$ m is considered.
The locations of upstream and downstream reservoirs are assumed to be $p_{\rm U}=0$ m and $p_{\rm D}=2000$ m, respectively. Two pressure sensors are situated at $x_1=1800$ m and $x_2=2000$ m. The wave speed is $a=1000$ m/s. The utilized frequencies are $w=jw_{th}$, $j=1,2,\ldots,32$, where $w_{th}=a\pi/(2l)$ is the fundamental frequency (first resonant frequency). Thus $N=64$. Under the hypothesis $H_1$, the leak location is $\phi=600$ m and the leak size is $s=1.4\times10^{-4}$ ${\rm m}^2$.
Other necessary parameters required in the system model (see Appendix \ref{appx:model}) are: $f=0.02$, $e^L=0$, $Q_0=0.0153$ ${\rm m}^3/{\rm s}$, $g=9.8$ ${\rm m}/{\rm s}^2$ and $H_{0}^L=23.5$ m. In the following simulations, we carry out Monte Carlo simulations using $10^5$ runs.


We compare the performance of our proposed LD-SCM scheme against alternative detection methods.  First, we consider the ``oracle" detector with perfect knowledge of parameters $s$, $\phi$ and ${\bf C}_N$. Although the oracle detector is unachievable in practice, it provides an upper bound on the performance of leak detection. We also compare with a classical method used in radar detection \cite{raghavan1995cfar}, which also uses the SCM as the estimate of ${\bf C}_N$ and is referred to as RD-SCM. Different from the LD-SCM scheme, this method estimates the leak component ${\bf p}=s{\bf g}(\phi)$ as a whole. It ignores the structure of ${\bf p}$ and does not estimate $s$ and $\phi$ separately.  Detailed descriptions of the oracle detector and the RD-SCM are provided in Appendix \ref{appx:benchmark}.

\begin{figure}[!htb]
\begin{center}
\subfigure[$P_{\rm D}$ against SNR with prescribed $P_{\rm FA}=10^{-3}$.]{
\includegraphics[width=0.9\linewidth]{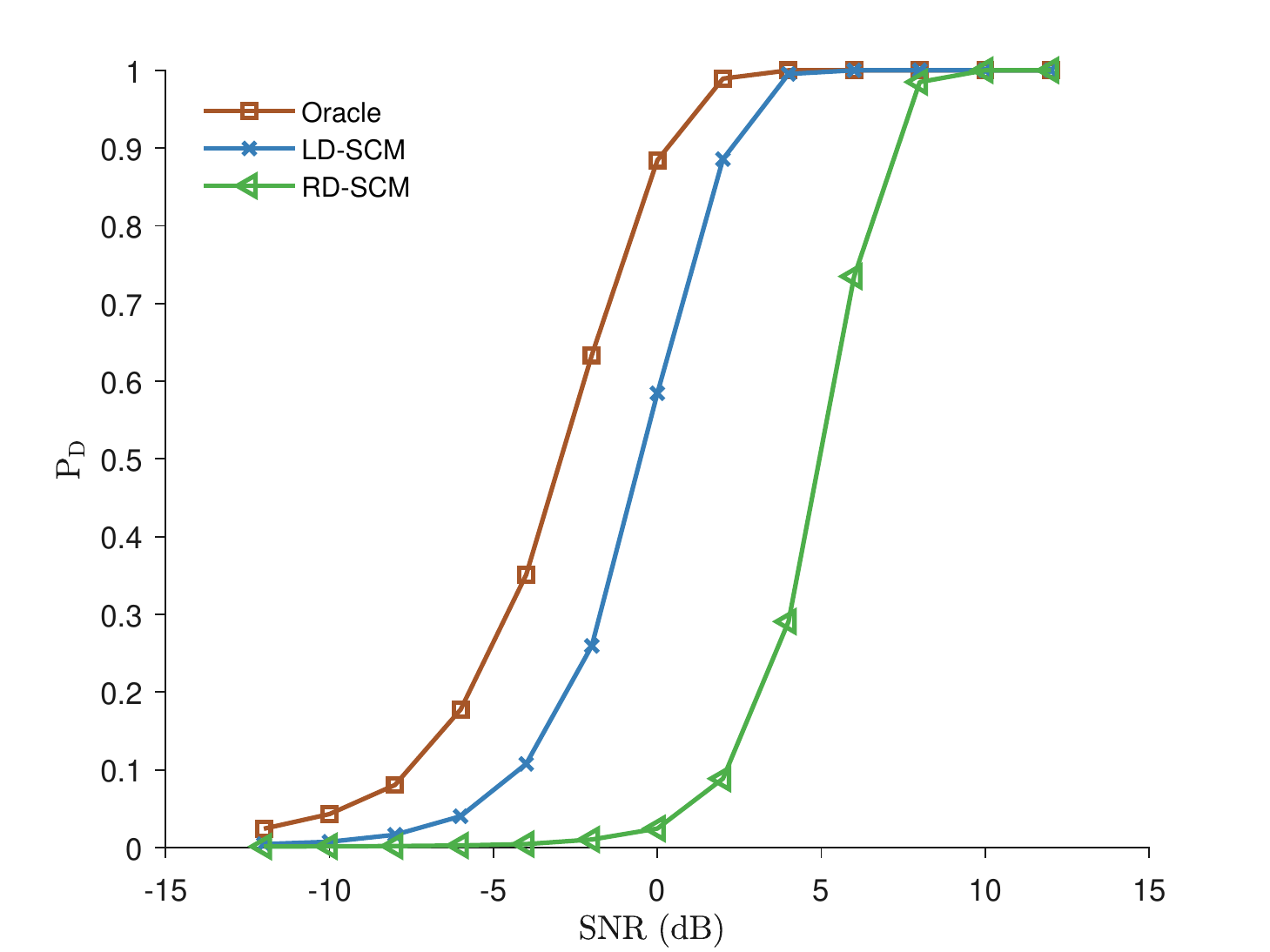}
\label{fig:SNR_Pd_3methods}}
\subfigure[ROCs with fixed SNR = -3 dB.]{
\includegraphics[width=0.9\linewidth]{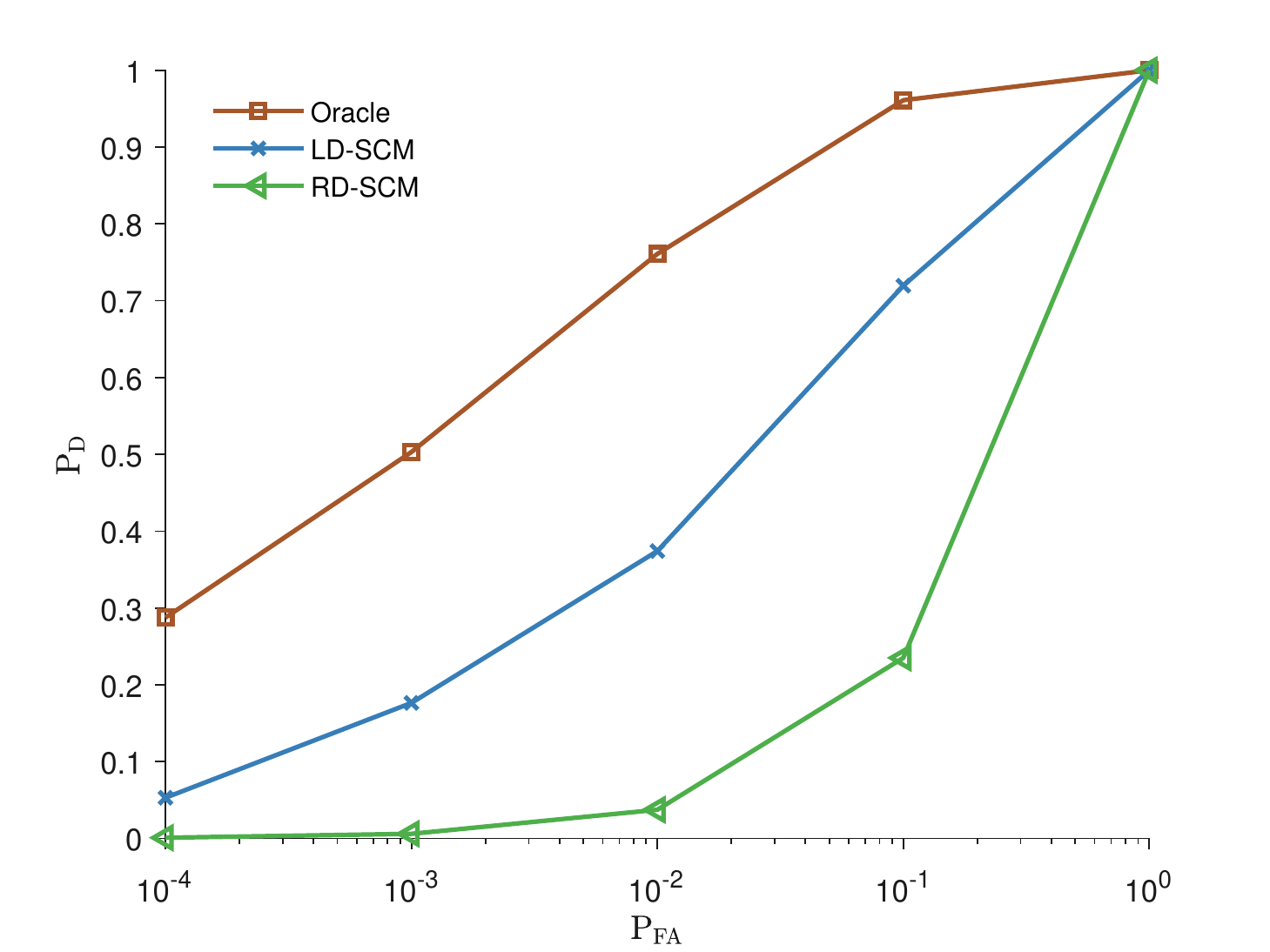}
\label{fig:ROC_3methods}}
\caption{Performance comparison of  the oracle detector, LD-SCM and RD-SCM when $N=64$, $K=600$.}
\label{fig:ROC_Pd_3methods}
\end{center}
\vspace{-0.3cm}
\end{figure}

In the simulations, we set $K=600$, $[{\bf C}_N]_{i,j}=\nu^20.9^{|i-j|}$ and define the signal to noise ratio (SNR) as ${\rm SNR}=\frac{\|{\bf p}\|^2}{\nu^2}$.
Fig.\;\ref{fig:SNR_Pd_3methods} shows the detection probability $P_{\rm D}$ against different SNRs under $P_{\rm FA}=10^{-3}$. Our proposed LD-SCM has higher $P_{\rm D}$ than that realized by the RD-SCM over different SNRs, and performs fairly close to the oracle.

To further demonstrate the performance of the LD-SCM, we plot receiver operating characteristic (ROC) curves for the different approaches. Fig.\;\ref{fig:ROC_3methods} shows that while the oracle detector naturally performs the best, the LD-SCM uniformly outperforms the RD-SCM over the entire span of $P_{\rm FA}$.


\begin{figure}[!htb]
\begin{center}
\includegraphics[width=0.9\linewidth]{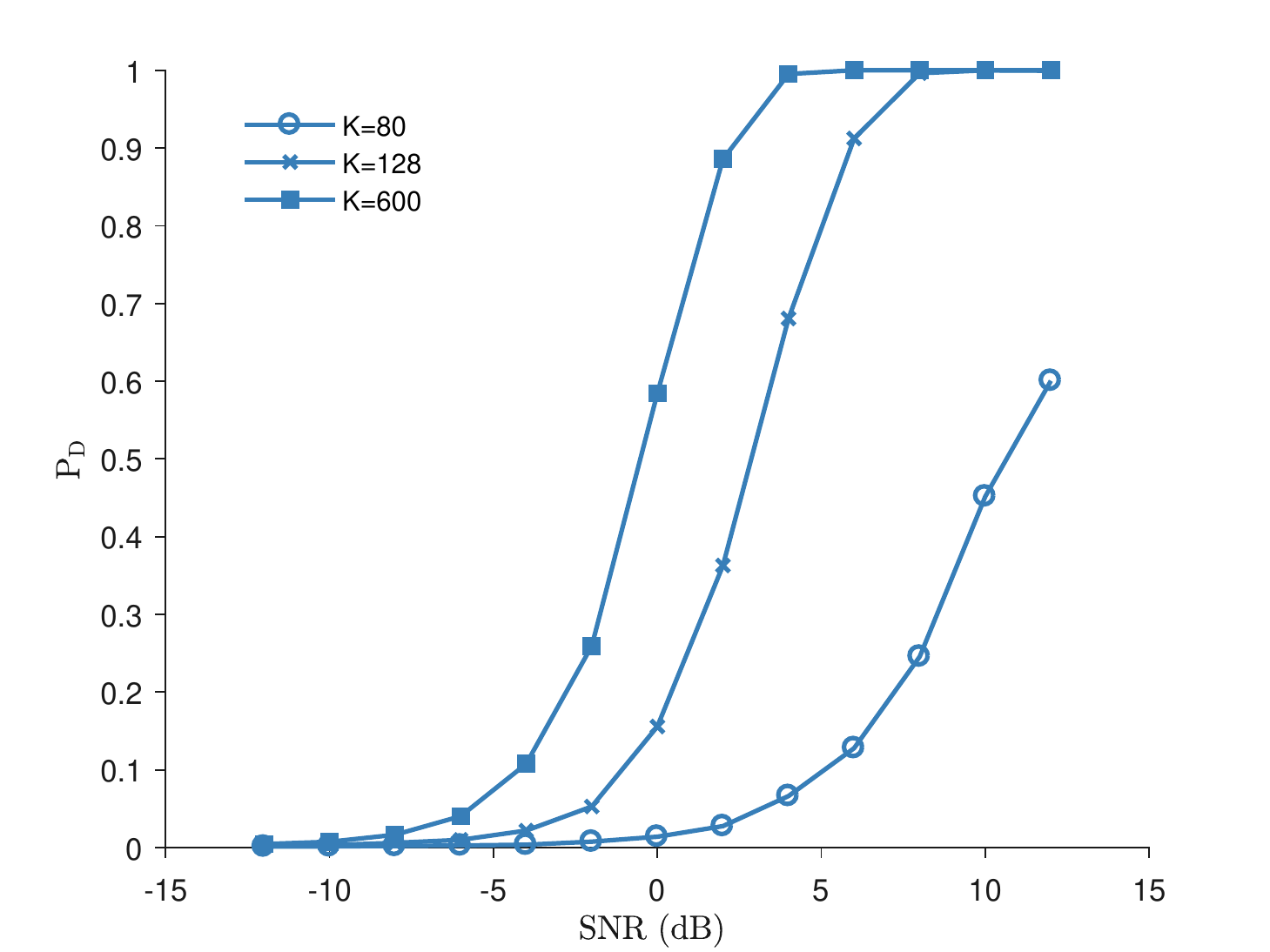}
\vspace{-0.2cm}
\caption{$P_{\rm D}$ of  LD-SCM against SNR with prescribed $P_{\rm FA}=10^{-3}$, for $N=64$ and different $K$.}
\label{fig:ROC_Pd_105}
\end{center}
\vspace{-0.3cm}
\end{figure}

To show the effect of the sample size $K$ of the secondary data, we further compare the leak detection performance of the LD-SCM for fixed $N=64$ and different $K$. As we see from Fig.\;\ref{fig:ROC_Pd_105}, the detection probability $P_{\rm D}$ decreases when $K$ becomes smaller. This is because the sample size $K$ is closely related to the estimation accuracy of the SCM.  It is well known that the estimation error of the SCM becomes large when the sample size $K$ is small compared to the data dimension $N$ \cite{reed1974rapid,boroson1980sample,mestre2008improved}. This has been demonstrated rigorously using random matrix theory, which considers the setting when $K$ and $N$ are both large, and which has shown that the eigenvalues and eigenvectors of the SCM behave very differently from those of ${\bf C}_N$ \cite{marvcenko1967distribution,mestre2008asymptotic,silverstein1995strong,silverstein1995empirical}. Thus, the performance degradation of the LD-SCM is caused in part by the estimation error of the SCM. To deal with this, a more robust covariance matrix estimate may help to enhance the leak detection performance when $K$ is not substantially larger than $N$. This is the main focus of the subsequent section.

\section{Leak detection with regularized sample covariance matrix}
As shown in the last section, the performance of the LD-SCM degrades when the sample size $K$ does not greatly exceed the matrix dimension $N$. Since the measurements are collected through $M$ sensors at $J$ frequencies, it is possible that the data dimension $N=J\times M$ is large, compared to the sample size $K$. Thus it is desirable to design a leak detection method that yields good performance when the data dimension is high or the sample size of the secondary data is small.  As the performance degradation is, to some extent, caused by the increased estimation error of the SCM, we may apply a more robust high dimensional covariance matrix estimator. A popular approach is the regularized SCM (RSCM) \cite{ledoit2004well,mestre2006finite,rubio2012performance,kammoun2018optimal}. We consider in this paper the design of an RSCM estimator, with the regularization parameter specifically optimized for the leak detection problem. We denote this second proposed leak detection scheme as LD-RSCM. It is inspired by recent works \cite{kammoun2018optimal,couillet2016second} on radar detection.

\subsection{Derivation of LD-RSCM  with unknown $\phi$ (under $H_1$)} \label{sec:knownPhi}
Initially, we introduce the design of the LD-RSCM with unknown leak location $\phi\in\mathcal{R}_l$ as $\mathcal{R}_l=[p_{\rm U}, p_{\rm D}]$ in the hypothesis $H_1$. The problem with unknown leak location (which is the case in practice) is addressed in Section \ref{sec:estphi}, in which the estimation of  $\phi$ under $H_1$ is considered.  With ${\bf g}(\phi)$ remained untouched, our data model becomes similar to that in radar detection \cite{kelly1986adaptive}. From results in \cite{kelly1986adaptive,gini1997sub}, the MLE of $s$ under $H_1$ is a function of $\phi$:
\begin{align}\label{eq:MLE_s}
\hat{s}(\phi)=\argmax_s f_1({\bf z}_0,\ldots,{\bf z}_K|H_1)=\frac{{\rm Re}\{{\bf g}^H(\phi){\bf C}_N^{-1}{\bf z}_0\}}{{\bf g}^H(\phi){\bf C}_N^{-1}{\bf g}(\phi)}.
\end{align}
By substituting $\hat{s}(\phi)$ for $s$ in $f_1$, the logarithm of the LR test statistic becomes
\begin{align}\label{L:knownC_N}
L(\phi)=\ln\frac{\max_{s}f_1({\bf z}_0,\ldots,{\bf z}_K|H_1)}{f_0({\bf z}_0,\ldots,{\bf z}_K|H_0)}=\frac{{\rm Re}^2\{{\bf g}^H(\phi){\bf C}_N^{-1}{\bf z}_0\}}{{\bf g}^H(\phi){\bf C}_N^{-1}{\bf g}(\phi)}.
\end{align}

Since ${\bf C}_N$ is unknown in (\ref{L:knownC_N}) and in order to cope with a possible deficiency in samples and improve the covariance matrix estimation accuracy, we use the RSCM as the estimate of ${\bf C}_N$, which is defined as follows:
\begin{align}\nonumber
\hat{\bf C}_N(\rho)=(1-\rho)\frac{N}{{\rm tr}({\bf R}_N)}{\bf R}_N+\rho{\bf I}_N,
\end{align}
where $\rho\in(0,1]$ is the regularization parameter and ${\bf R}_N=\frac{1}{K}\sum_{k=1}^K{\bf z}_k{\bf z}_k^H$ is the SCM computed with the secondary data. We normalize the trace of ${\bf R}_N$ to be of the same scale with that of ${\bf I}_N$ to ensure $\hat{\bf C}_N(\rho)$ to be sensitive to $\rho$. By plugging the RSCM into the test statistic $L(\phi)$ in (\ref{L:knownC_N}), we obtain $L(\rho,\phi)$ as a function of the regularization parameter $\rho$ and the leak location $\phi$, and the hypothesis test becomes
\begin{align}\label{eq:L_rho}
L(\rho,\phi)=\frac{{\rm Re}^2\{{\bf g}^H(\phi)\hat{\bf C}_N^{-1}(\rho){\bf z}_0\}}{{\bf g}^H(\phi)\hat{\bf C}_N^{-1}(\rho){\bf g}(\phi)}\mathop{\gtrless}^{H_1}_{H_0}\alpha.
\end{align}
Our aim is to find the optimal $\rho$, for any $\phi$ (which would be estimated), that can asymptotically maximize the detection probability $P_{\rm D}=\mathbb{P}\left[L(\rho,\phi)>\alpha|H_1\right]$ under a pre-determined false alarm probability $P_{\rm FA}=\mathbb{P}\left[L(\rho,\phi)>\alpha|H_0\right]$. For fixed $N$ and $K$, this is not an easy task. Additionally, it is obvious to see that the distribution of $L(\rho,\phi)$ in (\ref{eq:L_rho}) depends on ${\bf C}_N$ and unlike the LD-SCM method, the LD-RSCM scheme does not enjoy the CFAR property. This adds to the difficulty of determining the threshold $\alpha$.

Inspired by \cite{couillet2016second,kammoun2018optimal}, we resort to asymptotic tools from random matrix theory to address this problem. The approach is to first characterize the asymptotic false alarm and detection probabilities for all $\rho$ within a specified range, under the assumption that $N,K\rightarrow\infty$ with $c_N=N/K\rightarrow c$. We subsequently provide consistent estimators of the asymptotic false alarm and detection probabilities that are defined only in terms of the observed primary and secondary data. Based on this, we fix the estimated false alarm probability and optimize online over $\rho$ to maximize the estimated detection probability.

Following this approach, we assume that $\limsup_N\|{\bf C}_N\|<\infty$ where $\|{\bf C}_N\|$ is the spectral norm of ${\bf C}_N$.
Additionally, we make an extra assumption on the order of magnitude of $s$ with respect to $N$ to avoid getting trivial limiting results as $N\rightarrow\infty$. To see this, consider hypothesis $H_1$, and recall (\ref{eq:H1H0}), noting that $\|{\bf g}(\phi)\|_2 = O(\sqrt{N})$ (since ${\bf g}(\phi)$ is an $N$-dimensional vector whose elements do not depend on $N$). Then, if $s$ remains fixed as $N \to \infty$, (\ref{eq:L_rho}) implies that $L(\rho,\phi)\rightarrow\infty$, and consequently, $P_{\rm D}\rightarrow1$ for any fixed threshold $\alpha$. In order to avoid this, we assume that $s = O(\frac{1}{\sqrt{N}})$. In practice, this indicates that a small leak size is considered, which makes the detection problem even more difficult.

 We first observe that the structure of $L(\rho,\phi)$ in (\ref{eq:L_rho}) is similar to that of the test statistic $\hat{T}_N^{\rm RSCM}(\rho)$ described in \cite{kammoun2018optimal}, which is
\begin{align}\nonumber
\hat{T}_N^{\rm RSCM}(\rho)=\frac{\left|{\bf g}^H\hat{\bf C}_N^{-1}(\rho){\bf z}_0\right|}{\sqrt{{\bf z}_0^H\hat{\bf C}_N^{-1}(\rho){\bf z}_0}\sqrt{{\bf g}^H\hat{\bf C}_N^{-1}(\rho){\bf g}}}.
\end{align}

The forms of $L(\rho,\phi)$ and $\hat{T}_N^{\rm RSCM}(\rho)$ are similar, but not exactly the same. Especially, in $\hat{T}_N^{\rm RSCM}(\rho)$, ${\bf g}$ is known, not parameterized by unknown $\phi$. Nonetheless, the subsequent analysis will draw significantly from the technical derivations in \cite{kammoun2018optimal} (also \cite{couillet2016second}).

To demonstrate our results, we first introduce some frequently used quantities.
Denote for $z\in\mathbb{C}\backslash\mathbb{R}_{+}$ by $m_N(z)$ the unique complex solution to
\small
\begin{align}\nonumber
m_N(z)\!=\!\left(\!-z\!+\!c_N(1-\rho)\frac{1}{N}{\rm tr}{\bf C}_N({\bf I}_N\!+\!(1\!-\!\rho)m_N(z){\bf C}_N)^{-1}\!\right)^{-1}\!\!\!.
\end{align}
\normalsize
Define for $\kappa>0$, $\mathcal{R}_{\kappa}$ as $\mathcal{R}_{\kappa}\triangleq[\kappa,1]$.
Also denote $\underline{\rho}\triangleq\frac{\rho}{\rho+\frac{(1-\rho)N}{{\rm tr}({\bf C}_N)}}$. With these notations at hand, we are now ready to analyze the asymptotic behaviors of $P_{\rm FA}$ and $P_{\rm D}$.
\begin{theorem}[False alarm probability]\label{th:Pfa}
Under the assumption that $\phi$ is independent of ${\bf z}_0, {\bf z}_1,\ldots,{\bf z}_K$, we have as $N, K\rightarrow\infty$, with $c_N=N/K\rightarrow c\in(0,1)$,
\begin{align}\nonumber
\sup_{\rho\in\mathcal{R}_{\kappa}, \phi\in\mathcal{R}_l}\left|\mathbb{P}\left[L(\rho,\phi)>\alpha|H_0\right]-Q_1\left(\frac{\alpha}{\sigma^2(\rho,\phi)}\right)\right|\rightarrow0,
\end{align}
where
\begin{align}\nonumber
&\sigma^2(\rho,\phi)=\frac{1}{2\rho}\frac{1}{{\bf g}^H(\phi){\bf Q}_N(\underline{\rho}){\bf g}(\phi)}\\ \label{eq:sigma}
&\hspace{1.2cm}\times\frac{{\bf g}^H(\phi){\bf C}_N{\bf Q}_N^2(\underline{\rho}){\bf g}(\phi)}{1-cm_N^2(-\underline{\rho})(1-\underline{\rho})^2\frac{1}{N}{\rm tr}{\bf C}_N^2{\bf Q}_N^2(\underline{\rho})}, \\ \nonumber
&{\bf Q}_N(\underline{\rho})=({\bf I}_N+(1-\underline{\rho})m_N(-\underline{\rho}){\bf C}_N)^{-1}
\end{align}
and $Q_1\left(\frac{\alpha}{\sigma^2(\rho,\phi)}\right)$ is the regularized gamma function\footnotemark[1]
\begin{align} \label{eq:Q1Defn}
Q_1\left(\frac{\alpha}{\sigma^2(\rho,\phi)}\right)=Q\left(\frac{1}{2}, \frac{\alpha}{2\sigma^2(\rho,\phi)}\right).
\end{align}
\end{theorem}
\footnotetext[1]{$Q(r,x)$ is defined as $Q(r,x)=\frac{\Gamma(r,x)}{\Gamma(r)}$
where the upper incomplete gamma function $\Gamma(r,x)$ is $\Gamma(r,x)=\int_{x}^\infty t^{r-1}e^{-t}dt$
and the gamma function $\Gamma(r)$ is $\Gamma(r)=\Gamma(r,0)=\int_0^\infty t^{r-1}e^{-t}dt$.
}
{\bf Proof:} See Appendix \ref{appx:th_Pfa}.

{This is a uniform convergence result over both $\rho$ and $\phi$, which is essential in the sequel. The uniform convergence over $\rho$ allows the design of setting $\rho$ that maximizes $P_{\rm D}$ at a certain $P_{\rm FA}$, while the uniform convergence over $\phi$ ensures Theorem\;\ref{th:Pfa} and the following results still hold with the unknown $\phi$ being replaced by its corresponding estimate.

The proof of Theorem \ref{th:Pfa} follows a similar methodology used in \cite{couillet2016second}. First, we prove the pointwise convergence for each $\rho\in\mathcal{R}_{\kappa}$ and $\phi\in\mathcal{R}_l$. Then we generalize the convergence result to uniform convergence across $\rho\in\mathcal{R}_{\kappa}$ and $\phi\in\mathcal{R}_l$. In contrast to \cite{couillet2016second}, the key challenge lies in the additional study of the uniform convergence across $\phi\in\mathcal{R}_l$. Due to the space limitation, detailed proof is included in the Supplementary Material S1.

Theorem\;\ref{th:Pfa} provides an asymptotic expression for $P_{\rm FA}$.
The following theorem provides an asymptotic expression for the detection probability $P_{\rm D}$.}

\begin{theorem}[Detection probability]\label{th:Pd}
Under the assumption that $\phi$ is independent of ${\bf z}_0, {\bf z}_1,\ldots,{\bf z}_K$, we have as $N, K\rightarrow\infty$, with $c_N\rightarrow c\in(0,1)$,
\begin{align}\nonumber
\sup_{\rho\in\mathcal{R}_{\kappa},\phi\in\mathcal{R}_l}\left|\mathbb{P}\left[L(\rho,\phi)>\alpha|H_1\right]-Q_2\left(\beta^2(\rho,\phi), \frac{\alpha}{\sigma^2(\rho,\phi)}\right)\right|\rightarrow0,
\end{align}
where $Q_2$ is
\begin{align}\nonumber
Q_2(\lambda,x)=e^{-\lambda/2}\sum_{j=0}^\infty\frac{(\lambda/2)^j}{j!}\frac{\gamma(\frac{1+2j}{2},x/2)}{\Gamma(\frac{1+2j}{2})}
\end{align}
while $\gamma(r,x)=\int_{0}^x t^{r-1}e^{-t}dt$,
and
\begin{align}\nonumber
\beta(\rho,\phi)=&\sqrt{2}s\frac{{\bf g}^H(\phi){\bf Q}_N(\underline{\rho}){\bf g}(\phi)}{\sqrt{{\bf g}^H(\phi){\bf C}_N{\bf Q}_N^2(\underline{\rho}){\bf g}(\phi)}}\\\nonumber
&\times\sqrt{1-cm_N^2(-\underline{\rho})(1-\underline{\rho})^2\frac{1}{N}{\rm tr}{\bf C}_N^2{\bf Q}_N^2(\underline{\rho})}.
\end{align}
\end{theorem}
{\bf Proof:} See Appendix \ref{appx:th_Pd}.

According to Theorem \ref{th:Pfa} and Theorem \ref{th:Pd}, $L(\rho,\phi)$ behaves quite differently depending on whether there is a leak in the water pipe or not. In particular, under $H_0$, $L(\rho,\phi)$ asymptotically behaves like a chi-squared random variable, with $1$ degree of freedom parameterized by $\sigma^2(\rho,\phi)$; while it is well approximated under $H_1$ by a noncentral chi-squared random variable with $1$ degree of freedom, parameterized by $\sigma^2(\rho,\phi)$ and $\beta^2(\rho,\phi)$.

We will now discuss the choice of the regularization parameter $\rho$ and the threshold $\alpha$. We aim at setting $\rho$ and $\alpha$ for any certain $\phi\in\mathcal{R}_l$ in such a way as to maximize the asymptotic $P_{\rm D}$, with the asymptotic $P_{\rm FA}$ set to a fixed (tolerable) value $\eta$. From Theorem \ref{th:Pfa}, one can easily see that the values of $\alpha$ and $\rho$ that provide an asymptotic $P_{\rm FA}$ equal to $\eta$ should satisfy
\begin{align}\nonumber
\frac{\alpha}{\sigma^2(\rho,\phi)}=Q_1^{-1}(\eta).
\end{align}
From these choices, we then look for those values that maximize the asymptotic detection probability which is given, according to Theorem \ref{th:Pd}, by
\begin{align}\nonumber
Q_2\left(\beta^2(\rho,\phi), \frac{\alpha}{\sigma^2(\rho,\phi)}\right).
\end{align}
The second argument of $Q_2$ should be kept fixed in order to ensure the required asymptotic $P_{\rm FA}$. Noting also that $Q_2$ increases with respect to the first argument, which depends on $\rho$ but not $\alpha$, the optimization of $P_{\rm D}$ boils down to considering any $\rho^*$ satisfying:
\begin{align}\label{eq:opt_rho}
\rho^*\in\argmax_{\rho\in\mathcal{R}_{\kappa}}\{\theta(\rho,\phi)\}
\end{align}
where $\theta(\rho,\phi)\triangleq\frac{1}{2s^2}\beta^2(\rho,\phi)$.
Note the presence of  ``$\in$" in (\ref{eq:opt_rho}), since the optimization on the right-hand side can adopt multiple solutions.
Then the corresponding threshold should be
\begin{align}\label{eq:alpha_opt}
\alpha^*=\sigma^2(\rho^*,\phi)Q_1^{-1}(\eta).
\end{align}
The maximal asymptotic $P_{\rm D}$ that can be obtained while satisfying an asymptotic $P_{\rm FA}$ equal to $\eta$ is thus given by
\begin{align}\nonumber
P_{\rm D}=Q_2\left(2s^2\theta(\rho^*,\phi), \frac{\alpha^*}{\sigma^2(\rho^*,\phi)}\right).
\end{align}

These solutions for $\rho^*$ and $\alpha^*$ should be seen as ``oracle'' solutions, since they are not directly realizable from measured data.  Specifically, they require knowledge of $\sigma^2(\rho,\phi)$ and $\theta(\rho,\phi)$, which involve the unknown covariance matrix ${\bf C}_N$ (and also the unknown $\phi$, to be addressed subsequently).  Hence, to provide a practically useful solution, it is necessary to obtain consistent estimates of $\sigma^2(\rho,\phi)$ and $\theta(\rho,\phi)$ based on the available sample data. Such estimates, which do not require specific knowledge of ${\bf C}_N$, are provided in the following propositions.

\begin{prop}\label{th:sigma}
For $\rho\in(0,1)$ and $\phi\in\mathcal{R}_l$, define
\begin{align}\label{eq:sigma_est}
\hat{\sigma}^2(\rho,\phi)=\frac{{\rm tr}({\bf R}_N)}{2(1-\rho)N}\frac{1-\frac{\rho{\bf g}^H(\phi)\hat{\bf C}_N^{-2}(\rho){\bf g}(\phi)}{{\bf g}^H(\phi)\hat{\bf C}_N^{-1}(\rho){\bf g}(\phi)}}{\left(1-c_N+c_N\rho\frac{1}{N}{\rm tr}\hat{\bf C}_N^{-1}(\rho)\right)^2}
\end{align}
and let $\hat{\sigma}^2(1,\phi)=\lim_{\rho\uparrow1}\hat{\sigma}^2(\rho,\phi)=\frac{{\bf g}^H(\phi){\bf R}_N{\bf g}(\phi)}{2{\bf g}^H(\phi){\bf g}(\phi)}$. Under the assumption that $\phi$ is independent of ${\bf z}_0, {\bf z}_1,\ldots,{\bf z}_K$, we have, as $N, K\rightarrow\infty$, with $c_N\rightarrow c\in(0,1)$,
\begin{align}\nonumber
\sup_{\rho\in\mathcal{R}_{\kappa},\phi\in\mathcal{R}_l}\left|\hat{\sigma}^2(\rho,\phi)-\sigma^2(\rho,\phi)\right|\stackrel{\rm a.s.}\longrightarrow0.
\end{align}
Moreover,
\begin{align}\nonumber
\sup_{\rho\in\mathcal{R}_{\kappa},\phi\in\mathcal{R}_l}\left|\mathbb{P}\left[L(\rho,\phi)>\alpha|H_0\right]-Q_1\left(\frac{\alpha}{\hat{\sigma}^2(\rho,\phi)}\right)\right|\rightarrow0.
\end{align}
\end{prop}
{\bf Proof:} See Appendix \ref{appx:th_sigma}.

\begin{prop}\label{th:theta_est}
For $\rho\in(0,1)$ and $\phi\in\mathcal{R}_l$, define $\hat{\theta}(\rho,\phi)$ as
\begin{align} \nonumber
\hat{\theta}(\rho,\phi)=&\frac{(1-\rho)N}{{\rm tr}({\bf R}_N)}\left(1-c_N+c_N\rho\frac{1}{N}{\rm tr}\hat{\bf C}_N^{-1}(\rho)\right)^2\\\label{eq:hatPhiDef}
&\times\frac{({\bf g}^H(\phi)\hat{\bf C}_N^{-1}(\rho){\bf g}(\phi))^2}{{\bf g}^H(\phi)\hat{\bf C}_N^{-1}(\rho){\bf g}(\phi)-\rho{\bf g}^H(\phi)\hat{\bf C}_N^{-2}(\rho){\bf g}(\phi)}
\end{align}
and let $\hat{\theta}(1,\phi)\triangleq\lim_{\rho\uparrow1}\hat{\theta}(\rho,\phi)=\frac{({\bf g}^H(\phi){\bf g}(\phi))^2}{{\bf g}^H(\phi){\bf S}_N{\bf g}(\phi)}$. Under the assumption that $\phi$ is independent of ${\bf z}_0, {\bf z}_1,\ldots,{\bf z}_K$, we have as $N, K\rightarrow\infty$, with $c_N\rightarrow c\in(0,1)$,
\begin{align}\nonumber
\sup_{\rho\in\mathcal{R}_{\kappa},\phi\in\mathcal{R}_l}\left|\hat{\theta}(\rho,\phi)-\theta(\rho,\phi)\right|\stackrel{\rm a.s.}\longrightarrow0.
\end{align}
Moreover
\small
\begin{align}\nonumber
\sup_{\rho\in\mathcal{R}_{\kappa},\phi\in\mathcal{R}_l}\left|\mathbb{P}\left[L(\rho,\phi)>\alpha|H_1\right]-Q_2\left(2{s}^2\hat{\theta}(\rho,\phi), \frac{\alpha}{\hat{\sigma}^2(\rho,\phi)}\right)\right|\rightarrow0.
\end{align}
\normalsize
\end{prop}
{\bf Proof:} Since the structure of $\theta(\rho,\phi)$ is similar to that of $\sigma^2(\rho,\phi)$, Proposition \ref{th:theta_est} can be proved similarly to Proposition \ref{th:sigma}.

Next, since the convergence results in Theorem \ref{th:Pd} and Proposition \ref{th:theta_est} are uniform in $\rho$, we can establish the following:
\begin{cor}\label{th:rho_est}
For $\phi\in\mathcal{R}_l$, define $\hat{\rho}^*$ as any value satisfying
\begin{align}\nonumber
\hat{\rho}^*\in\argmax_{\rho\in\mathcal{R}_{\kappa}}\hat{\theta}(\rho,\phi).
\end{align}
Under the assumption that $\phi$ is independent of ${\bf z}_0, {\bf z}_1,\ldots,{\bf z}_K$, for every $\alpha>0$ and $\phi\in\mathcal{R}_l$, as $N,K\rightarrow\infty$ with $c_N\rightarrow c\in(0,1)$,
\begin{align}\nonumber
\left|\mathbb{P}\left[L(\hat{\rho}^*,\phi)>\alpha|H_1\right]-\max_{\rho\in\mathcal{R}_{\kappa}}\{\mathbb{P}\left[L(\rho,\phi)>\alpha|H_1\right]\}\right|\rightarrow0.
\end{align}
\end{cor}
{\bf Proof:} This can be proved following the same steps as in the proof of \cite[Corollary 1]{couillet2016second}, and therefore is omitted.

Hence, $\hat{\rho}^*$ provides an asymptotically optimal estimate of $\rho^*$.  Moreover, from (\ref{eq:alpha_opt}) and Proposition \ref{th:sigma}, we construct a consistent estimate of ${\alpha}^*$ (for achieving an asymptotic $P_{\rm FA}$ of a prescribed value $\eta$) as follows:
\begin{align}\nonumber
\hat{\alpha}=\hat{\sigma}^2(\hat{\rho}^*,\phi)Q_1^{-1}(\eta).
\end{align}

The final remaining issue, required to establish a completely data-dependent leak detection algorithm, is to address the problem of unknown $\phi$.  This is pursued in the following.

%

\subsection{Estimation of unknown leak location $\phi$}\label{sec:estphi}

Here we develop an estimator $\hat{\phi}$ and correspondingly ${\bf g}(\hat{\phi})$ that can be substituted for the unknown ${\bf g}(\phi)$ in the test statistic $L(\rho,\phi)$ in (\ref{eq:L_rho}).
From (\ref{eq:pdf_f1}) and (\ref{eq:MLE_s}), the MLE of $\phi$ with measurement ${\bf z}_0$ is
\begin{align}\nonumber
\hat{\phi}&=\argmax_{\phi\in[p_{\rm U}, p_{\rm D}]}f_1({\bf z}_0, {\bf z}_1, \ldots,{\bf z}_K|{\bf C}_N, \hat{s}, H_1)\\\label{eq:Est_phi_CN}
&=\argmax_{\phi\in[p_{\rm U}, p_{\rm D}]}\frac{{\rm Re}^2\{{\bf g}^H(\phi){\bf C}_N^{-1}{\bf z}_0\}}{{\bf g}^H(\phi){\bf C}_N^{-1}{\bf g}(\phi)}.
\end{align}
However, the MLE of $\hat{\phi}$ in (\ref{eq:Est_phi_CN}) is based on the unobservable ${\bf C}_N$. In the following theorem, we show that the estimate $\hat{\phi}_{\{{\bf R}_N,{\bf z}_0\}}$ given by (\ref{eq:Est_phi_CN}) but with ${\bf C}_N$ replaced by the SCM ${\bf R}_N$, is asymptotically equivalent to the estimate $\hat{\phi}$ in (\ref{eq:Est_phi_CN}).

\begin{theorem}\label{th:phi_est}
Define $\hat{\phi}_{\{{\bf R}_N,{\bf z}_0\}}$ as any value satisfying
\begin{align}\label{eq:phi_scm}
\hat{\phi}_{\{{\bf R}_N,{\bf z}_0\}}\in\argmax_{\phi\in\mathcal{R}_l}\frac{{\rm Re}^2\{{\bf g}^H(\phi){\bf R}_N^{-1}{\bf z}_0\}}{{\bf g}^H(\phi){\bf R}_N^{-1}{\bf g}(\phi)}.
\end{align}
As $N, K\rightarrow\infty$, with $c_N=N/K\rightarrow c\in(0,1)$,
\begin{align}\nonumber
\!\!\left|\frac{{\rm Re}^2\{{\bf g}^H(\hat{\phi}_{\{{\bf R}_N,{\bf z}_0\}}){\bf C}_N^{-1}{\bf z}_0\}}{{\bf g}^H(\hat{\phi}_{\{{\bf R}_N,{\bf z}_0\}}){\bf C}_N^{-1}{\bf g}(\hat{\phi}_{\{{\bf R}_N,{\bf z}_0\}})}\!-\!\frac{{\rm Re}^2\{{\bf g}^H(\hat{\phi}){\bf C}_N^{-1}{\bf z}_0\}}{{\bf g}^H(\hat{\phi}){\bf C}_N^{-1}{\bf g}(\hat{\phi})}\right|\stackrel{\rm a.s.}\longrightarrow0.
\end{align}
\end{theorem}
{\bf Proof}: See Appendix \ref{appx:th_phi}.

With $\hat{\phi}_{\{{\bf R}_N,{\bf z}_0\}}$, and correspondingly ${\bf g}(\hat{\phi}_{\{{\bf R}_N,{\bf z}_0\}})$, the test statistic $L(\rho,\phi)$ in (\ref{eq:L_rho}) becomes, by substituting ${\bf g}(\hat{\phi}_{\{{\bf R}_N,{\bf z}_0\}})$ for ${\bf g}(\phi)$,
\begin{align}\nonumber
{L}(\hat{\phi}_{\{{\bf R}_N,{\bf z}_0\}},\rho)=\frac{{\rm Re}^2\{{\bf g}^H(\hat{\phi}_{\{{\bf R}_N,{\bf z}_0\}})\hat{\bf C}_N^{-1}(\rho){\bf z}_0\}}{{\bf g}^H(\hat{\phi}_{\{{\bf R}_N,{\bf z}_0\}})\hat{\bf C}_N^{-1}(\rho){\bf g}(\hat{\phi}_{\{{\bf R}_N,{\bf z}_0\}})}.
\end{align}
However, it is difficult to study the asymptotic $P_{\rm FA}$ and $P_{\rm D}$ of statistic ${L}(\rho,\hat{\phi}_{\{{\bf R}_N,{\bf z}_0\}})$, unlike the analysis of ${L}(\rho,\phi)$ given in Theorem \ref{th:Pfa} and Theorem \ref{th:Pd}, in which $\phi$ is independent of ${\bf z}_0, {\bf z}_1, \ldots, {\bf z}_K$. As we can see from (\ref{eq:phi_scm}), $\hat{\phi}_{\{{\bf R}_N,{\bf z}_0\}}$ depends on primary data ${\bf z}_0$ and ${\bf R}_N$ constructed from the secondary data ${\bf z}_1,\ldots,{\bf z}_k$. This dependency makes the asymptotic analysis of ${L}(\rho,\hat{\phi}_{\rho,\{{\bf R}_N,{\bf z}_0\}})$ even more complicated.

If we were to have access to a parallel independent set of data for estimating $\phi$ (i.e., ${\bf y}_0$ in place of ${\bf z}_0$, and ${\bf y}_1, \ldots, {\bf y}_K$ in place of ${\bf z}_1, \ldots, {\bf z}_K$), such that
\begin{align} \label{eq:phiHatIdeal}
\hat{\phi}_{\{{\bf W}_N,{\bf y}_0\}}\in\argmax_{\phi\in\mathcal{R}_l}\frac{{\rm Re}^2\{{\bf g}^H(\phi){\bf W}_N^{-1}{\bf y}_0\}}{{\bf g}^H(\phi){\bf W}_N^{-1}{\bf g}(\phi)}
\end{align}
where ${\bf W}_N=\frac{1}{K}\sum_{k=1}^K{\bf y}_k{\bf y}_k^H$, then $\hat{\phi}_{\{{\bf W}_N,{\bf y}_0\}}$ is independent of ${\bf z}_0, {\bf z}_1, \ldots, {\bf z}_K$ and all the results presented in Section\;\ref{sec:knownPhi} hold upon substituting ${\bf g}(\hat{\phi}_{\{{\bf W}_N,{\bf y}_0\}})$ for ${\bf g}(\phi)$.

In the absence of such parallel data set, however, we can still apply the proposed statistic ${L}(\rho,\phi)$, but it will generally be suboptimal. Nonetheless, through simulations, which are not shown due to space limitations, we find that in practice there is no need to have a complete parallel data set to achieve good performance, but rather, it is sufficient to simply have access to ${\bf y}_0$. This is because the correlations induced by using ${\bf z}_1, \ldots, {\bf z}_K$ in estimating $\phi$ are rather weak and thus minimally affect  performance, whereas the dependencies induced by ${\bf z}_0$ are strong and lead to substantial performance degradation. Thus, we propose to employ the estimator $\hat{\phi}_{\{{\bf R}_N,{\bf y}_0\}}$ as any value satisfying
\begin{align} \label{eq:phiHatFinal}
\hat{\phi}_{\{{\bf R}_N,{\bf y}_0\}}\in\argmax_{\phi\in[p_{\rm U}, p_{\rm D}]}\frac{{\rm Re}^2\{{\bf g}^H(\phi){\bf R}_N^{-1}{\bf y}_0\}}{{\bf g}^H(\phi){\bf R}_N^{-1}{\bf g}(\phi)}.
\end{align}
Based on the results in Section \ref{sec:knownPhi} with the estimated leak location $\hat{\phi}_{\{{\bf R}_N,{\bf y}_0\}}$ substituted for $\phi$, we obtain the optimized regularization parameter $\hat{\rho}^*$ and test statistic ${L}(\hat{\rho}^*,\hat{\phi}_{\{{\bf R}_N,{\bf y}_0\}})$. {Both $\hat{\phi}_{\{{\bf R}_N,{\bf y}_0\}}$ and $\hat{\rho}^*$ can be computed through simple numerical searches in the range of $\mathcal{R}_l$ and $\mathcal{R}_{\kappa}$ respectively.} Our proposed leak detection scheme, LD-RSCM, is summarized in {\bf Algorithm \ref{al:alg2}}.

\begin{algorithm}[h]
{\small{
\caption{LD-RSCM}\label{al:alg2}
\begin{enumerate}
\smallskip
\item Compute the estimated leak location $\hat{\phi}_{\{{\bf R}_N,{\bf y}_0\}}$ based on (\ref{eq:phiHatFinal}).
\item Set the regularization parameter $\hat\rho^*$ as
\begin{align}\nonumber
\hat{\rho}^*\in\argmax_{\rho\in\mathcal{R}_{\kappa}}\hat{\theta}(\rho,\hat{\phi}_{\{{\bf R}_N,{\bf y}_0\}})
\end{align}
with $\hat{\theta}(\cdot)$ given by (\ref{eq:hatPhiDef}), but with $\phi$ replaced by $\hat{\phi}_{\{{\bf R}_N,{\bf y}_0\}}$.

\item For a user-prescribed false alarm probability $\eta$, set the threshold $\hat{\alpha}$ as
\begin{align}\nonumber
\hat{\alpha}=\hat{\sigma}^2(\hat{\rho}^*,\hat{\phi}_{\{{\bf R}_N,{\bf y}_0\}})Q_1^{-1}(\eta)
\end{align}
with $Q_1(\cdot)$ defined in (\ref{eq:Q1Defn}) and $\hat{\sigma}(\cdot)$ defined as in (\ref{eq:sigma_est}), but with $\phi$ replaced by $\hat{\phi}_{\{{\bf R}_N,{\bf y}_0\}}$.

\item Construct the test statistic
\begin{align}\nonumber
{L}(\hat{\rho}^*,\hat{\phi}_{\{{\bf R}_N,{\bf y}_0\}})=\frac{{\rm Re}^2\{{\bf g}^H(\hat{\phi}_{\{{\bf R}_N,{\bf y}_0\}})\hat{\bf C}_N^{-1}(\hat{\rho}^*){\bf z}_0\}}{{\bf g}^H(\hat{\phi}_{\{{\bf R}_N,{\bf y}_0\}})\hat{\bf C}_N^{-1}(\hat{\rho}^*){\bf g}(\hat{\phi}_{\{{\bf R}_N,{\bf y}_0\}})}.
\end{align}
\item Accept $H_0$ (``no leak"), if $\tilde{L}(\hat{\rho}^*,\hat{\phi}_{\{{\bf R}_N,{\bf y}_0\}})\leq\hat{\alpha}$; otherwise accept $H_1$ (``leak present").
\item If $H_1$ accepted, set the estimates of $\phi$ and $s$:
\begin{align}\nonumber
\hat{\phi}=\hat{\phi}_{\{{\bf R}_N,{\bf y}_0\}},~~
\hat{s}=\frac{{\rm Re}\{{\bf g}^H(\hat{\phi}_{\{{\bf R}_N,{\bf y}_0\}})\hat{\bf C}_N^{-1}(\hat{\rho}^*){\bf z}_0\}}{{\bf g}^H(\hat{\phi}_{\{{\bf R}_N,{\bf y}_0\}})\hat{\bf C}_N^{-1}(\hat{\rho}^*){\bf g}(\hat{\phi}_{\{{\bf R}_N,{\bf y}_0\}})}.
\end{align}
\end{enumerate}
}}
\end{algorithm}

\subsection{Simulation Results}\label{sec:simulation}
Here we present simulation results to test the performance of the proposed leak detection algorithm, LD-RSCM. We consider a scenario with $K$ comparable to $N$, setting $K=128$, $N=64$. Other than the choice of $K$ and $N$, the same simulation settings are used as described in Section \ref{sec:sim_3methods}. Results are averaged over Monte Carlo simulations of $10^5$ runs.

{\subsubsection{Accuracy of theoretical approximations for false alarm and detection probabilities}

We start by checking the accuracy of the asymptotic theoretical results for the false alarm probability.  Specifically, for ${L}(\rho,\phi)$ in (\ref{eq:L_rho}), in Fig.\;\ref{fig:Theorem1_H0} we plot the exact value of $P_{\rm FA} = \mathbb{P}\left[L(\rho,\phi)>\alpha|H_0\right]$ (computed empirically), and compare with the deterministic asymptotic approximation $Q_1\left( \alpha / \sigma^2(\rho,\phi)\right)$ from Theorem \ref{th:Pfa}, and the corresponding approximation with estimated $\hat{\sigma}^2(\rho,\phi)$, $Q_1\left( \alpha / \hat{\sigma}^2(\rho,\phi) \right)$ from Proposition \ref{th:sigma}. All curves are in good agreement.  We further check the accuracy of the asymptotic theoretical results for the detection probability in Fig.\;\ref{fig:Theorem2_H1_v2}, plotting the exact value of $P_{\rm D} = \mathbb{P}\left[L(\rho,\phi)>\alpha|H_1\right]$ (computed empirically), along with the deterministic asymptotic  approximation $Q_2\left(\beta^2(\rho,\phi), \frac{\alpha}{\sigma^2(\rho,\phi)}\right)$ from Theorem \ref{th:Pd}, and the corresponding approximation with estimated values of $\beta^2(\rho,\phi)$ and $\sigma^2(\rho,\phi)$, $Q_2\left(2{s}^2\hat{\theta}(\rho,\phi), \frac{\alpha}{\hat{\sigma}^2(\rho,\phi)}\right)$, from Proposition \ref{th:theta_est}. Again, we see close alignment between the theoretical and empirical results.
\begin{figure}[!htb]
\begin{center}
\subfigure[$P_{\rm FA}$, empirical and theoretical.]{
\includegraphics[width=0.8\linewidth]{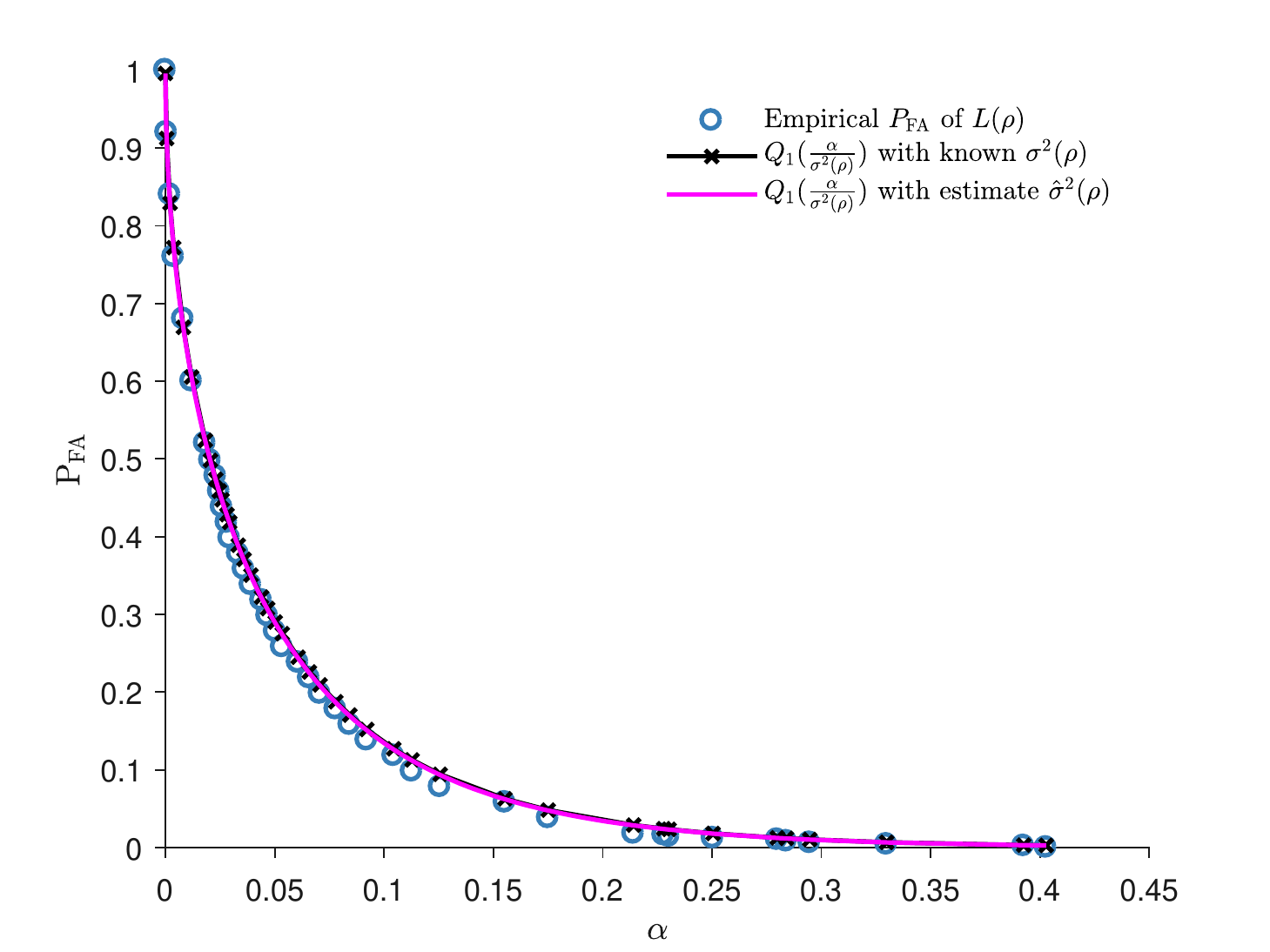}
\label{fig:Theorem1_H0}}
\subfigure[$P_{\rm D}$, empirical and theoretical. Results for $P_{\rm FA}=10^{-3}$.]{
\includegraphics[width=0.8\linewidth]{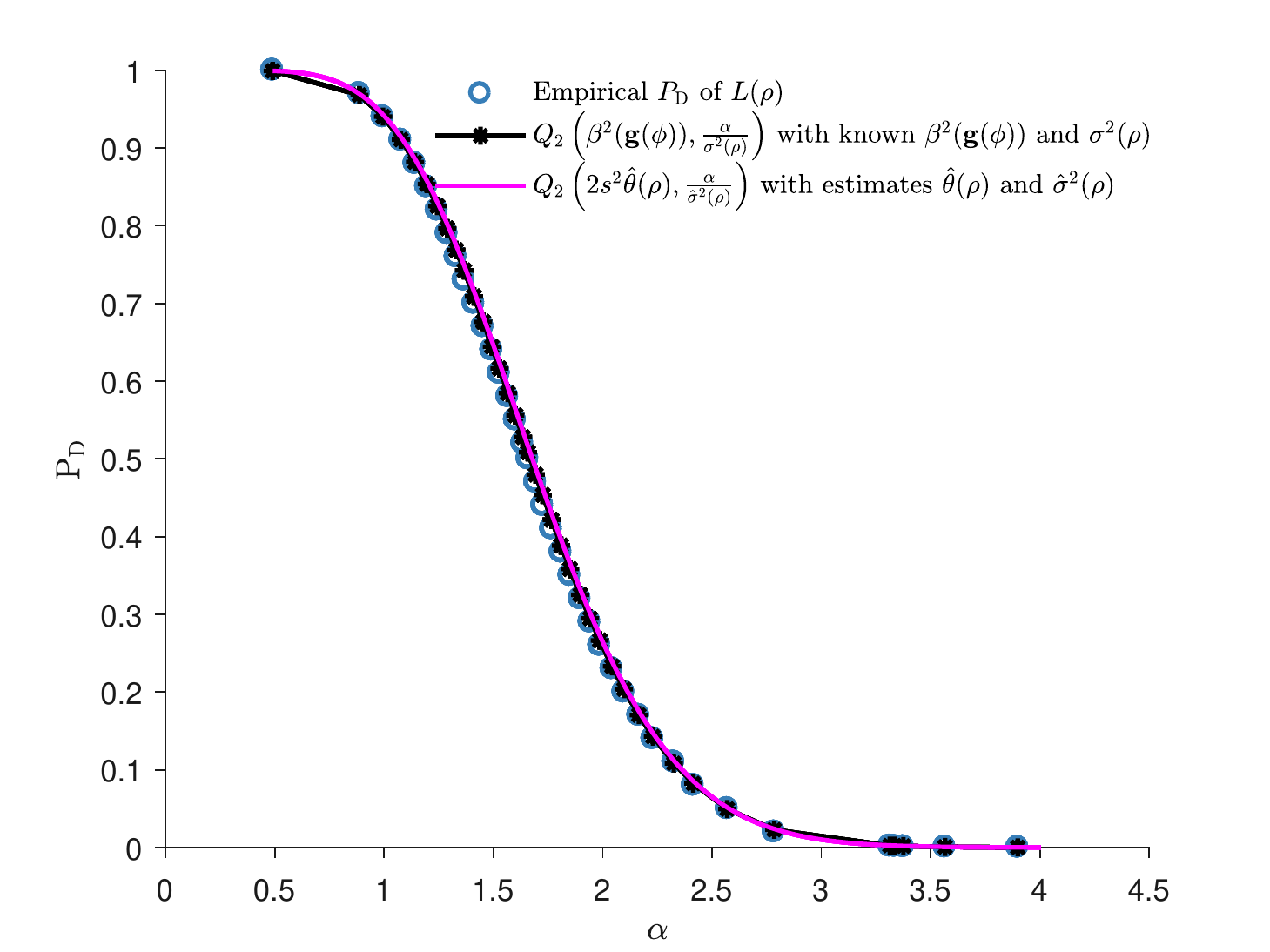}
\label{fig:Theorem2_H1_v2}}
\caption{Empirical and theoretical results for the false alarm and detection probabilities achieved with ${L}(\rho)$. Results for $[{\bf C}_N]_{i,j}=\nu^20.9^{|i-j|}$, ${\rm SNR} =  -3$ dB, $\phi=600$ m and $\rho=0.5$. }
\label{fig:Theorem_knownphi}
\end{center}
\end{figure}

\subsubsection{Performance of the proposed test statistic with different $\phi$ estimators}
Next we check the performance, in terms of both false alarm probability and detection probability, of the proposed test statistic ${L}(\rho,\hat{\phi})$ when constructed from different estimates of $\phi$.  Specifically, in Fig.\;\ref{fig:Theorem_diffphi}, we compare $P_{\rm FA}$ and $P_{\rm D}$ (computed empirically) for ${L}(\rho,\hat{\phi})$ constructed using $\hat{\phi}_{\{{\bf W}_N,{\bf y}_0\}}$, $\hat{\phi}_{\{{\bf R}_N,{\bf y}_0\}}$ and $\hat{\phi}_{\{{\bf R}_N,{\bf z}_0\}}$, with ${\bf W}_N$ and ${\bf y}_0$ defined as in (\ref{eq:phiHatIdeal}).  We first observe that if $\phi$ is estimated using ${\bf R}_N$ and ${\bf z}_0$ (equivalently, from ${\bf z}_0, {\bf z}_1, \ldots, {\bf z}_K$), the performance deteriorates substantially, at least in terms of false alarm probability.  On the other hand, the performance is similar whether $\phi$ is estimated based on ${\bf R}_N$ and ${\bf y}_0$ or from ${\bf W}_N$ and ${\bf y}_0$, confirming the claims made above, leading to the proposed estimate in (\ref{eq:phiHatFinal}).  Moreover, as shown in the figure, even though not theoretically concrete, our asymptotic  approximations for the false alarm and detection probabilities remain accurate for $\phi$ estimates constructed from ${\bf R}_N$ and ${\bf y}_0$, but they completely break down when such estimates are constructed from ${\bf R}_N$ and ${\bf z}_0$.  This reinforces the need for the additional independent sample ${\bf y}_0$, for the proposed algorithm to perform well.
\begin{figure}[!htb]
\begin{center}
\subfigure[$P_{\rm FA}$, empirical and theoretical]{
\includegraphics[width=0.85\linewidth]{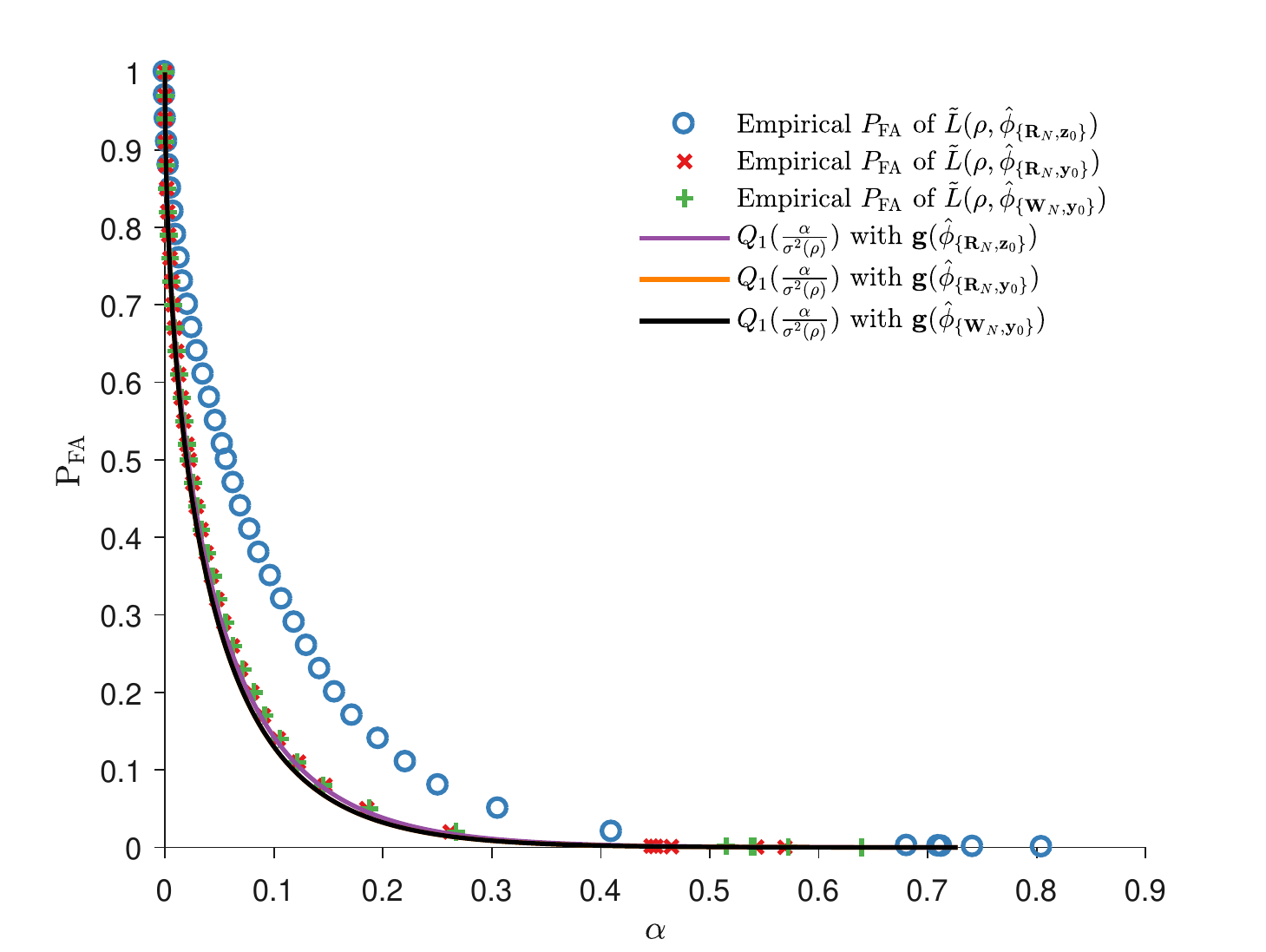}
\label{fig:Theorem1_H0_diffphi}}
\subfigure[$P_{\rm D}$, empirical and theoretical. Results for $P_{\rm FA}=10^{-3}$]{
\includegraphics[width=0.85\linewidth]{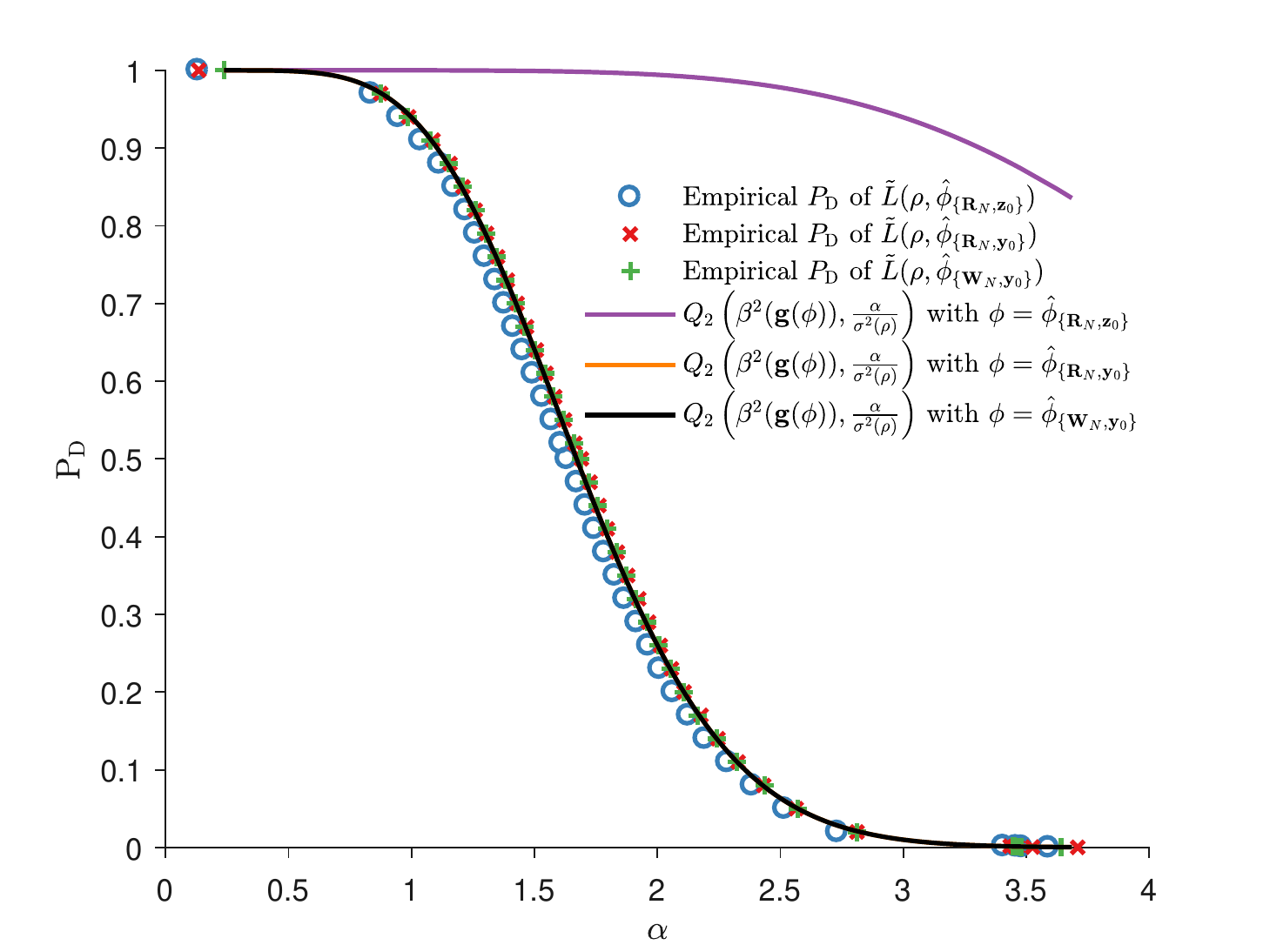}
\label{fig:Theorem2_H1_diffphi}}
\caption{Empirical and theoretical results for the false alarm and detection probabilities achieved with $L(\hat{\phi},\rho)$, with different estimators $\hat{\phi}$. Results for $[{\bf C}_N]_{i,j}=\nu^20.9^{|i-j|}$, ${\rm SNR}=-3$ dB, and $\rho=0.5$. }
\label{fig:Theorem_diffphi}
\end{center}
\vspace{-0.3cm}
\end{figure}
}

\subsubsection{Performance comparison of LD-RSCM and LD-SCM}
\begin{figure}[!htb]
\begin{center}
\subfigure[$P_{\rm D}$ against SNR with prescribed $P_{\rm FA}=10^{-3}$.]{
\includegraphics[width=0.9\linewidth]{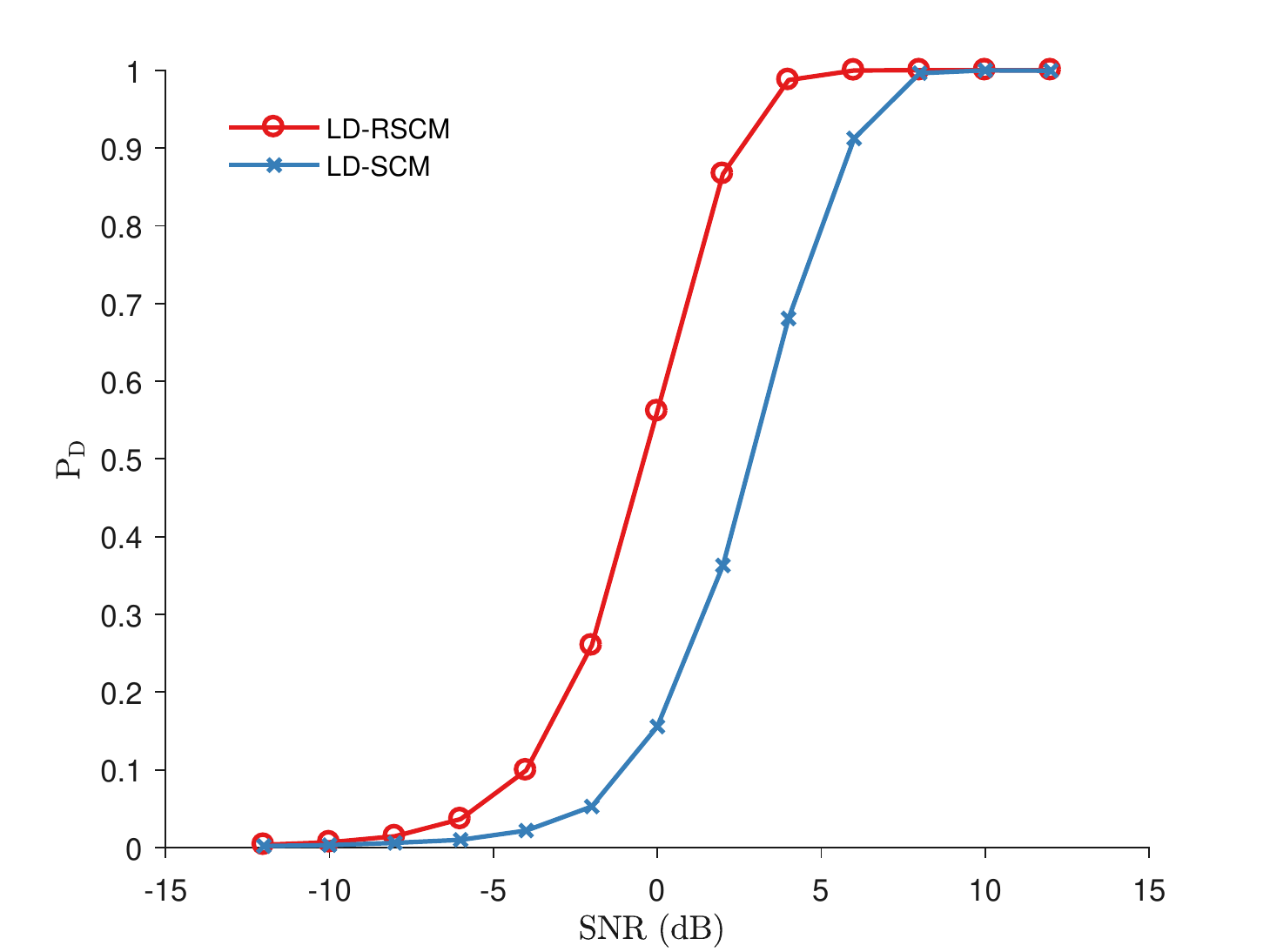}
\label{fig:SNR_Pd_Pfa3}}
\subfigure[ROCs with fixed SNR = -3 dB.]{
\includegraphics[width=0.9\linewidth]{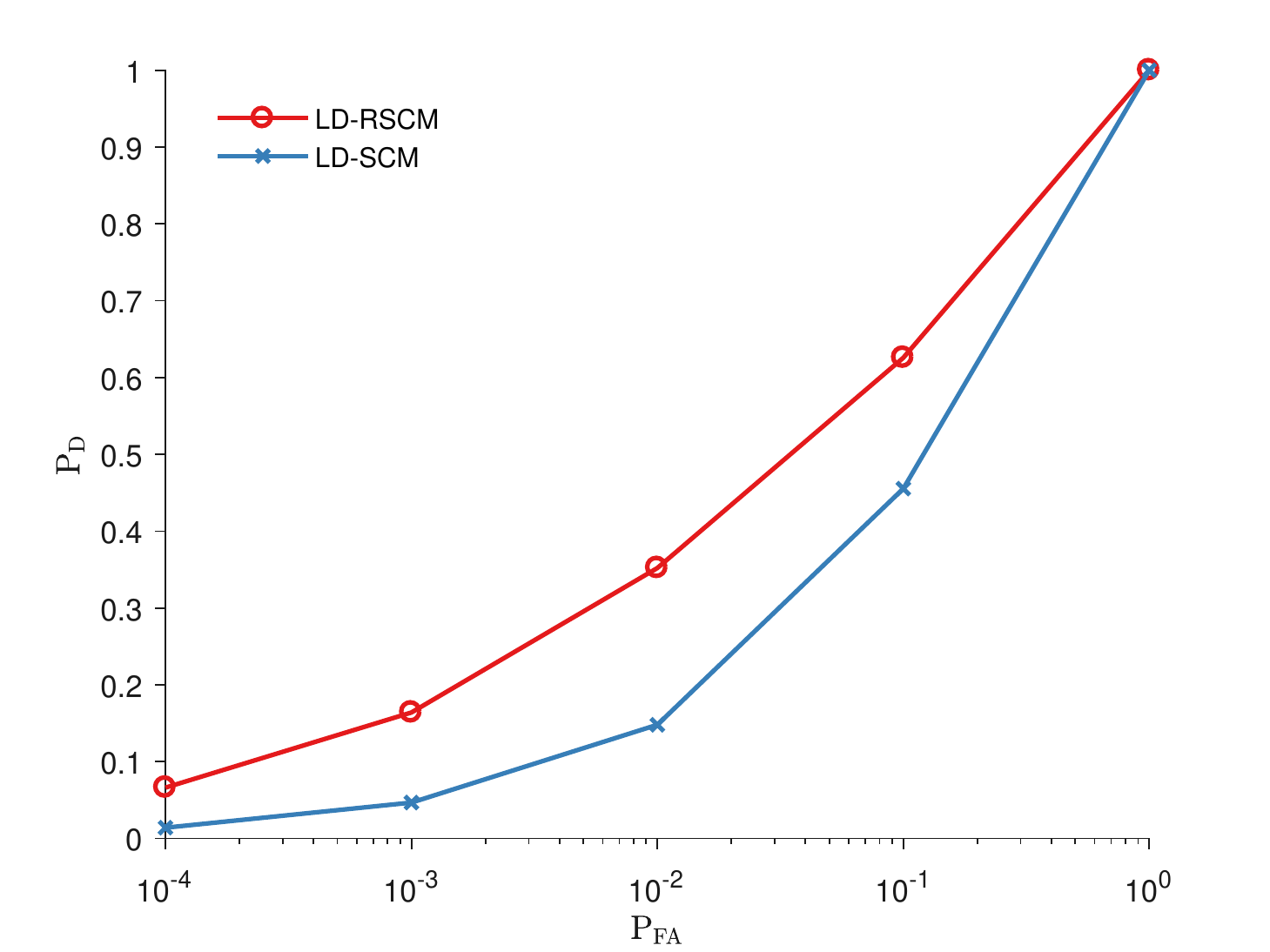}
\label{fig:ROC_semilogx}}
\caption{Performance comparison of LD-RSCM and LD-SCM when $N=64$, $K=128$.}
\label{fig:ROC_Pd_128}
\end{center}
\end{figure}

We compute the performance of the proposed LD-RSCM leak detector, and compare this against the LD-SCM detector that we proposed earlier.  For the implementation of LD-RSCM, we assume having an extra primary data ${\bf y}_0$, which is not needed in LD-SCM. In
Fig.\;\ref{fig:SNR_Pd_Pfa3} we plot the detection probability $P_{\rm D}$ against SNR, for $P_{\rm FA}=10^{-3}$, $[{\bf C}_N]_{i,j}=\nu^20.9^{|i-j|}$. Evidently, LD-RSCM achieves higher detection probability than LD-SCM over the entire span of SNRs. Performance gains are also reflected in Fig.\;\ref{fig:ROC_semilogx}, which presents ROC curves for ${\rm SNR}=-3$ dB.  These results clearly demonstrate the advantage of employing a robust covariance matrix estimate to achieve superior leak detection accuracy under high dimensional settings.  

%

\section{Discussion}
This paper has presented methods for automatically detecting leaks in a water pipeline. This is an important problem for practical water supply systems, which are plagued by inefficiencies caused by pipeline leakages. Such leakages can not only lead to loss of valuable natural resources, but they can also lead to compromised water quality and potentially affect public health.

As we have shown, the leak detection problem naturally can be formulated as a binary hypothesis test which, technically, amounts to detecting structured signals (originating due to leakages) in the presence of correlated noise.  By adopting the GLRT testing principle, we proposed a simple test procedure which we demonstrated to perform well, particularly when the number of measured samples is not low. The proposed method also has the practically-desirable CFAR property.  To further improve performance under data limited (or high-dimensional) scenarios, we further leveraged results from random matrix theory to present a more robust solution.  This method revealed better performance, at the expense of requiring higher implementation complexity.

Overall, our work provides a first attempt at designing hypothesis tests which are specifically tailored for the problem of detecting leaks in pipelines.  Further experimental work will be needed to confirm the performance of the methods in the field. Moreover, an important extension will be to generalize the framework, possibly using multiple hypothesis testing theory, to detect multiple leaks in a pipeline, and to handle more complex pipeline configurations.

\begin{appendices}
{\section{Water pipeline signal model description}\label{appx:model}
Here we provide a brief introduction for the physical model in Section \ref{sec:model}, considering a water pipeline with a single leak. Especially we give a discussion about the derivations of $h_m^0(w_j)$ and $g_m(\phi, w_j)$ in the model. Further discussion about the model can be found in \cite{wang2018pipeline,wylie1993fluid,chaudhry1979applied}.

The discharge and head oscillations due to a fluid transient are represented by $q$ and $h$.
These are described by the linearized unsteady-oscillatory continuity and momentum equations in the time domain  \cite{chaudhry1979applied}
\begin{equation}\label{continuity_time}
\frac{\partial q}{\partial x}+\frac{gA}{a^2}\frac{\partial h}{\partial t}-\frac{Q_0^L}{2(H_0^L-e^L)}h(\phi)\delta(x-\phi)=0 ,
\end{equation}
\begin{equation}\label{momentum_time}
\frac{1}{gA}\frac{\partial q}{\partial t}+\frac{\partial h}{\partial x}+Rq=0,
\end{equation}
for $x\in[p_{\rm U},p_{\rm D}]$, in which $a$ is the wave speed, $g$ is the gravitational acceleration, $A$ is the area of the pipeline, $\phi$ is the leak location, $Q_0^L$ and $H_0^L$ are the steady-state discharge and head at the leak, $e^L$ is the elevation of the pipe at the leak, $R$ is the steady-state resistance term being $R=(fQ_0)/(gDA^2)$ for turbulent flows, $f$ is the Darcy-Weisbach friction factor, $Q_0$ is the steady-state discharge in the pipe and $D$ is the pipe diameter.
Physically, \eqref{continuity_time} represents the mass conservation principle. The first term in the left hand side of \eqref{continuity_time} is the divergence of mass at a point $x$ along the pipe. The second term represents the rate of accumulation of mass at $x$. Therefore, a net mass flux towards $x\neq\phi$ (i.e., $\frac{a^2}{gA}\frac{\partial q}{\partial x}<0$) is accommodated by mass accumulation towards $x$ (i.e., $\frac{\partial h}{\partial t}>0$). This accumulation is fundamentally due to the compressibility of the fluid and the elasticity of the pipe.
The last term in the left hand side of \eqref{continuity_time} depicts the mass conservation at the leak.
Let $\phi^{-}$ and $\phi^{+}$ represent respectively just upstream and just downstream of the leak.
With the assumption
\begin{equation}\label{eq:h at leak}
h(\phi^{-})=h(\phi^{+})=h(\phi),
\end{equation}
Eq.~\eqref{continuity_time} leads to
\begin{equation}\label{eq:q_leak}
q(\phi^{-})=q(\phi^{+})+q(\phi)=q(\phi^{+})-\frac{Q_0^L}{2(H_0^L-e^L)}h(\phi).
\end{equation}
Eq.~\eqref{momentum_time} is Newton's second law along the pipe. The first term ($\frac{1}{gA}\frac{\partial q}{\partial t}$) has its origin in the axial acceleration of the fluid. The second term ($\frac{\partial h}{\partial x}$) represents the net pressure force. The third term ($Rq$) is the resistance force due to the friction between the fluid and pipe wall.

Readers with electrical engineering background should note that there is a one to one correspondence between \eqref{continuity_time} and \eqref{momentum_time} and the Telegrapher equations \cite{pozar2009microwave}. The head $h$ is analogous to the voltage; the flow rate of fluid $q$ is analogous to the current; the friction coefficient $R$ is analogous to the resistance; $\frac{gA}{a^2}$ is the capacitance; $\frac{1}{gA}$ is the inductance; $\frac{Q_0^L}{2(H_0^L-e^L)}$ is analogous to the conductance of the shunt.

The model in this paper considers momentum along the pipe, but neglects momentum in the radial and azimuthal directions. This implies that the current model is for low frequency waves where the wavelength is much larger than the pipe diameter. In addition, the model is linearized  (i.e., nonlinear terms are neglected). This assumption is valid if (i) the wave amplitude is much lower than the steady-state pressure and (ii) the Mach number $\ll 1$. Typically, the steady-state pressure head is in the range 40 m to 70 m. Therefore, the assumption (i) is not limiting in practice. In addition, in practice the flow velocity is of the order of 1 m/s and the wave speed range is from 350 m/s to 1500 m/s. Therefore, the Mach number is of order 1/350 or less. Thus, the assumption (ii) is also not of concern in practice.

Taking the Fourier transform of~\eqref{continuity_time} and (\ref{momentum_time}) with respect to $t$ gives $q$ and $h$ in the frequency domain for $x\in[p_{\rm U},\phi)\cup(\phi,p_{\rm D}]$:
\begin{equation}\label{continuity_frequency}
\frac{a^2}{gA}\frac{\partial q}{\partial x}+{\rm i}wh=0 ,
\end{equation}
\begin{equation}\label{momentum_frequency}
\frac{\partial h}{\partial x}+\left(\frac{{\rm i}w}{gA}+R\right)q=0,
\end{equation}
where $w$ is the angular frequency. Solving~\eqref{continuity_frequency} and \eqref{momentum_frequency} with the head and mass conservation conditions across the leak, i.e., \eqref{eq:q_leak} and \eqref{eq:h at leak},
the quantities at $x_m$ can be computed in the following way \cite{chaudhry1979applied}:
\begin{small}
\begin{equation}\label{eq:transfer}
\left(\!\!\begin{array}{c}q(x_m)\\h(x_m)\end{array}\!\!\right)\!=\!M_0(x_m-\phi)\!\left(\!\!\begin{array}{cc}1 & -\frac{Q_0^L}{2(H_0^L-e^L)}\\0 & 1\end{array}\!\!\right)\!M_0(\phi)\!\left(\!\!\begin{array}{c}q(p_{\rm U})\\h(p_{\rm U})\end{array}\!\!\right)\!.
\end{equation}
\end{small}
In this equation,
\begin{equation}\label{matrix_noleak}
M_0(x)=\left(\begin{array}{cc} \cosh\left(\mu x\right) & -\frac{1}{Z}\sinh\left(\mu x\right)\\ -Z\sinh\left(\mu x\right) & \cosh\left(\mu x\right) \end{array}\right)
\end{equation}
is the field matrix, where $Z=\mu a^2/({\rm i}w g A)$ is the characteristic impedance and $\mu=a^{-1}\sqrt{-w^2+{\rm i}gAw R}$ is the propagation function.
If the pipe is frictionless ($f=0$), $\mu={\rm i}k$, where $k=w/a$ is the wavenumber.

The transfer matrix on the right hand side of~\eqref{eq:transfer} can be simplified as \cite{wang2018pipeline}:
\begin{equation}\label{eq:transfer2}
M_0(x_m-\phi)\left(\hspace{-0.2cm}\begin{array}{cc}1 & -\frac{Q_0^L}{2(H_0^L-e^L)}\\0 & 1\end{array}\hspace{-0.2cm}\right)M_0(\phi)=M_0(x_m)+sM_1(\phi),
\end{equation}
in which
\small
\begin{align}\nonumber
&M_1(\phi)=\sqrt{\frac{g}{2(H_0^L-e^L)}}\\\label{matrix^Leak}
&\!\!\times\!\!\left(\hspace{-0.3cm}\begin{array}{cc} Z\sinh\left(\mu \phi\right)\cosh\left(\mu(x_m\!-\!\phi)\right) &\hspace{-0.3cm}-\cosh\left(\mu \phi\right)\cosh\left(\mu(x_m\!-\!\phi)\right)\\ -Z^2\sinh\left(\mu \phi\right)\sinh\left(\mu(x_m\!-\!\phi)\right) &\hspace{-0.3cm}Z\cosh\left(\mu \phi\right)\sinh\left(\mu(x_m\!-\!\phi)\right) \end{array}\!\!\!\!\right)
\end{align}
\normalsize
is a matrix related to the location $\phi$ of the leak but independent of the leak size $s$.

By combining~\eqref{eq:transfer}--\eqref{matrix^Leak}, the head at
$x_m$ for a given angular frequency $w_j$ is
\begin{equation}\nonumber
h_m(w_j)=h^0_m(w_j)+sg_m(\phi,w_j),
\end{equation}
wherein
\begin{align}\nonumber
h^0_m(w_j)=&-Z(w_j)\sinh\left(\mu(w_j) x_m\right)q(p_{\rm U},w_j) + \\\nonumber
&\cosh\left(\mu(w_j) x_m\right)h(p_{\rm U},w_j)
\end{align}
and
\begin{align}\nonumber
&g_m(\phi,w_j)=-\frac{\sqrt{g}Z(w_j)\sinh(\mu(w_j)(x_m-\phi))}{\sqrt{2(H_0^L-e^L)}}\\\nonumber
&\!\!\times\!\left(Z(w_j)\sinh(\mu(w_j) \phi)q(p_{\rm U}, w_j)\!-\!\cosh(\mu(w_j)\phi)h(p_{\rm U},w_j)\right).
\end{align}
Applying the boundary condition that $h(p_{\rm U},w_j)=0$ (as the upstream $p_{\rm U}$ is connected to a reservoir), then
\begin{equation}\nonumber
h^0_m(w_j)=-Z(w_j)\sinh\left(\mu(w_j) x_m\right)q(p_{\rm U},w_j)
\end{equation}
and
\begin{align}\nonumber
g_m(\phi,w_j)=&-\frac{\sqrt{g}Z(w_j)\sinh(\mu(w_j)(x_m-\phi))}{\sqrt{2(H_0^L-e^L)}}\\\nonumber
&\times Z(w_j)\sinh(\mu(w_j) \phi)q(p_{\rm U}, w_j),
\end{align}
where $q(p_{\rm U},\omega_j)$ can be estimated by \cite{wang2018identification}
\begin{equation}\nonumber
q(p_{\rm U},\omega_j)=-\frac{h(p_{\rm U}+\epsilon, w_j)}{Z(\omega_j)\sinh(\mu(\omega_j)\epsilon)},
\end{equation}
where $h(p_{\rm U}+\epsilon, w_j)$ is a pressure head measured at a location very close to $p_{\rm U}$ (denoted by $p_{\rm U}+\epsilon$ where $0<\epsilon\ll l$).

Since the measured head $h_m(w_j)$ is contaminated by noise $n_m(w_j)$, it can be represented as
\begin{equation}\nonumber
h_m(w_j)=h^{0}_m(w_j)+sg_m(\phi, w_j)+n_m(w_j).
\end{equation}
}

\section{Benchmark methods}\label{appx:benchmark}
\subsection{Oracle detector}\label{appx:opt}
If for benchmarking purposes one supposes that under hypothesis $H_1$ the leak size $s$, leak location $\phi$ and noise covariance matrix ${\bf C}_N$ are assumed known, then the likelihood ratio test can be applied (instead of the GLRT), which maximizes the detection probability $P_{\rm D}$ at a certain false alarm probability $P_{\rm FA}$ \cite{lehmann2006testing}. For this oracle detector, from (\ref{eq:pdf_f0}) and (\ref{eq:pdf_f1}), the logarithm of the likelihood ratio statistic is equal to
\small
\begin{align}\nonumber
L&=\ln\frac{f_1({\bf z}_0,\ldots,{\bf z}_K)}{f_0({\bf z}_0,\ldots,{\bf z}_K)}\\\nonumber
&=2{\rm Re}\{s{\bf g}^H(\phi){\bf C}_N^{-1}{\bf z}_0\}-s{\bf g}^H(\phi){\bf C}_N^{-1}s{\bf g}(\phi).
\end{align}
\normalsize
Comparing with a threshold $\alpha$ results in the following optimal decision rule:
\begin{align}\nonumber
2{\rm Re}\{s{\bf g}^H(\phi){\bf C}_N^{-1}{\bf z}_0\}-s{\bf g}^H(\phi){\bf C}_N^{-1}s{\bf g}(\phi)\mathop{\gtrless}^{H_1}_{H_0}\alpha,
\end{align}
which, after straightforward simplification, can be rewritten as
\begin{align}\nonumber
\Delta_{\rm oracle}={\rm Re}\{s{\bf g}^H(\phi){\bf C}_N^{-1}{\bf z}_0\}\mathop{\gtrless}^{H_1}_{H_0}\alpha_2
\end{align}
where $\alpha_2=\frac{1}{2}(\alpha+s{\bf g}^H(\phi){\bf C}_N^{-1}s{\bf g}(\phi))$.  With this statistic, the false alarm probability is given by \mbox{$P_{\rm FA}=P[\Delta_{\rm oracle}>\alpha_2|H_0]$}, and the detection probability is given by $P_{\rm D}=P[\Delta_{\rm oracle}>\alpha_2|H_1]$.

It is important to note that the assumption of $s$, $\phi$, and ${\bf C}_N$ being known is not practically meaningful, but nonetheless, this oracle detector provides an upper bound on the performance that can be achieved by GLRT-based methods, which estimate these unknown quantities.

\subsection{RD-SCM}\label{appx:RD_SCM}
In our data model described in Section \ref{sec:model}, the leak component ${\bf p}$ is parameterized by the unknown leak size $s$ and the leak location $\phi$. If we were to ignore the structure of ${\bf p}$ and estimate this vector as a whole, the solution of the resulting leak detection problem would be the same as that considered previously in radar detection \cite{raghavan1995cfar}. We refer to this method as RD-SCM, as indicated in Section \ref{sec:sim_3methods}. In this case, the GLRT becomes:
\begin{align}\nonumber
L_2=\frac{\max_{{\bf C}_N}\max_{\bf p}f_1({\bf z}_0,\ldots,{\bf z}_K)}{\max_{{\bf C}_N}f_0({\bf z}_0,\ldots,{\bf z}_K)}\mathop{\gtrless}^{H_1}_{H_0}\alpha.
\end{align}
Under $H_0$, the MLE of ${\bf C}_N$ is $\frac{1}{K+1}\sum_{k=0}^K{\bf z}_k{\bf z}_k^H$, whereas under $H_1$, the MLEs of ${\bf p}$ and ${\bf C}_N$  are ${\bf z}_0$ and $\frac{1}{K+1}\sum_{k=1}^K{\bf z}_k{\bf z}_k^H$ respectively \cite{raghavan1995cfar}. Thus
\small
\begin{align}\nonumber
L_2=\left(\frac{\det\left({\bf z}_0{\bf z}_0^H+\sum_{k=1}^K{\bf z}_k{\bf z}_k^H\right)}{\det\left(\sum_{k=1}^K{\bf z}_k{\bf z}_k^H\right)}\right)^{K+1}.
\end{align}
\normalsize
Denote ${\bf S}_N=\sum_{k=1}^K{\bf z}_k{\bf z}_k^H$ and since
\small
\begin{align}
\det\left({\bf z}_0{\bf z}_0^H+\sum_{k=1}^K{\bf z}_k{\bf z}_k^H\right)=\det\left({\bf S}_N\right)\left(1+{\bf z}_0^H{\bf S}_N^{-1}{\bf z}_0\right),
\end{align}
\normalsize
the GLRT becomes
\begin{align}\label{eq:GLRT_1}
L_2=\left(1+{\bf z}_0^H{\bf S}_N^{-1}{\bf z}_0\right)^{K+1}\mathop{\gtrless}^{H_1}_{H_0}\alpha.
\end{align}
the GLRT (\ref{eq:GLRT_1}) is equivalent to the following test:
\begin{align}\nonumber
\Delta_2={\bf z}_0^H{\bf S}_N^{-1}{\bf z}_0\mathop{\gtrless}^{H_1}_{H_0}\alpha_3
\end{align}
where $\alpha_3=\sqrt[K+1]{\alpha}-1$.

One advantage of this approach is that the probability densities of $\Delta_2$ under $H_0$ and $H_1$ can be obtained analytically, as given in \cite{raghavan1995cfar,shah1998performance}. Thus, $P_{\rm FA}$ and $P_{\rm D}$ for this RD-SCM scheme can be written in closed-form \cite{raghavan1995cfar,shah1998performance}. We can also observe that the probability distribution of $\Delta_2$ is independent of ${\bf C}_N$ under $H_0$, and thus the RD-SCM also has the CFAR property, which is illustrated in detail in \cite{raghavan1995cfar}.

\section{Technical proofs}
\subsection{Proof of Theorem \ref{th:Pfa}}\label{appx:th_Pfa}
The proof follows by applying the methodology used in \cite{couillet2016second}. First, we prove the convergence for each $\rho\in\mathcal{R}_{\kappa}$ and $\phi\in\mathcal{R}_l$. We characterize the asymptotic behavior of the denominator and numerator of $L(\rho,\phi)$ separately.
Shown in \cite{kammoun2018optimal},  as $N,K\rightarrow\infty$, with $c_N=N/K\rightarrow c\in(0,1)$, the following results hold:
\begin{align}\label{eq:conv_gphi}
\left|\frac{1}{N}{\bf g}^H(\phi)\hat{\bf C}_N^{-1}(\rho){\bf g}(\phi)-\frac{1}{N\rho}{\bf g}^H(\phi){\bf Q}_N(\underline{\rho}){\bf g}(\phi)\right|\stackrel{\rm a.s.}\longrightarrow 0
\end{align}
and for some $x\sim N(0,1)$,
\small
\begin{align}\nonumber
&\frac{1}{\sqrt{N}}{\rm Re}({\bf g}^H(\phi)\hat{\bf C}_N^{-1}(\rho){\bf z}_0)-\\\nonumber
&\sqrt{\frac{1}{2\rho^2N}\frac{{\bf g}^H(\phi){\bf C}_N{\bf Q}_N^2(\underline{\rho}){\bf g}(\phi)}{1-cm_N(-\underline{\rho})^2(1-\underline{\rho})^2\frac{1}{N}{\rm tr}{\bf C}_N^2{\bf Q}_N^2(\underline{\rho})}}x=o_{p}(1)
\end{align}
\normalsize

This shows in particular that \mbox{$\frac{1}{N}{\rm Re}^2\{{\bf g}^H(\phi)\hat{\bf C}_N^{-1}(\rho){\bf z}_0\}$} behaves asymptotically as a chi-squared random variable with \mbox{scale $\sqrt{\frac{1}{2\rho^2N}\frac{{\bf g}^H(\phi){\bf C}_N{\bf Q}_N^2(\underline{\rho}){\bf g}(\phi)}{1-cm_N(-\underline{\rho})^2(1-\underline{\rho})^2\frac{1}{N}{\rm tr}{\bf C}_N^2{\bf Q}_N^2(\underline{\rho})}}$} and degree of freedom $1$. Using this result along with Slutsky's lemma \cite{gut2013probability}, we conclude that, under $H_0$, $L(\rho,\phi)$ is also asymptotically equivalent to a chi-squared random variable but with scale $\sigma(\rho,\phi)$. We therefore get, for fixed $\rho\in\mathcal{R}_{\kappa}$ and $\phi\in\mathcal{R}_l$,
\begin{align}\label{eq:result1}
\left|\mathbb{P}\left[L(\rho,\phi)>\alpha|H_0\right]-Q_1\left(\frac{\alpha}{\sigma^2(\rho,\phi)}\right)\right|\rightarrow0.
\end{align}
$Q_1\left(\frac{\alpha}{\sigma^2(\rho,\phi)}\right)$ is the regularized gamma function\footnotemark[1]
\begin{align} \label{eq:Q1Defn}
Q_1\left(\frac{\alpha}{\sigma^2(\rho,\phi)}\right)=Q\left(\frac{1}{2}, \frac{\alpha}{2\sigma^2(\rho,\phi)}\right).
\end{align}
The generalization to uniform convergence across $\rho\in\mathcal{R}_{\kappa}$ then follows via the same arguments as in \cite{couillet2016second}.

Next we prove the uniform convergence across $\phi\in\mathcal{R}_l$. 
To reduce the amount of notations, we drop the parameter $\rho$ in function $L(\rho,\phi)$ and covariance estimator ${\bf C}_N(\rho,\phi)$ in the following. We shall exploit a $\phi$-Lipschitz property of $L(\phi)$ to reduce the uniform convergence over $\mathcal{R}_l$ to a uniform convergence over finitely many values of $\phi$.

The $\phi$-Lipschitz property we shall need is as follows: for each $\varepsilon>0$,
\begin{align}\label{eq:lipschitz}
\lim_{\delta\rightarrow0}\lim_{N\rightarrow\infty}P\left(\sup_{\phi,\phi'\in\mathcal{R}_l\atop|\phi-\phi'|<\delta}|L(\phi)-L(\phi')|>\varepsilon\right)=0.
\end{align}
Let us prove this result. Let $\eta>0$ be small and $\mathcal{A}_N^\eta\triangleq\{\exists\phi\in\mathcal{R}_l,\frac{1}{N}{\bf g}^H(\phi)\hat{\bf C}_N^{-1}{\bf g}(\phi)<\eta\}$. Developing the difference $L(\phi)-L(\phi')$ and isolating the denominator according to its belonging to $\mathcal{A}_N^\eta$ or not, we may write
\begin{align}\nonumber
&P\left(\sup_{\phi,\phi'\in\mathcal{R}_l\atop|\phi-\phi'|<\delta}|L(\phi)-L(\phi')|>\varepsilon\right) \\ \nonumber
&\leq P(\mathcal{A}_N^\eta)+P\left(\sup_{\phi,\phi'\in\mathcal{R}_l\atop|\phi-\phi'|<\delta}V_N(\phi,\phi')>\varepsilon\eta\right)
\end{align}
where
\begin{align}\nonumber
V_N(\phi,\phi')\triangleq&\frac{1}{N^2}{\rm Re}^2\{{\bf g}^H(\phi)\hat{\bf C}_N^{-1}{\bf z}_0\}{\bf g}^H(\phi')\hat{\bf C}_N^{-1}{\bf g}(\phi') \\ \nonumber
&-\frac{1}{N^2}{\rm Re}^2\{{\bf g}^H(\phi')\hat{\bf C}_N^{-1}{\bf z}_0\}{\bf g}^H(\phi)\hat{\bf C}_N^{-1}{\bf g}(\phi).
\end{align}

It is obvious that $P(\mathcal{A}_N^\eta)\rightarrow0$ for a sufficiently small choice of $\eta$. To prove that
\begin{align}\nonumber
\lim_{\delta\rightarrow0}\limsup_{N}P\left(\sup_{|\phi-\phi'|<\eta}V_N(\phi,\phi')>\varepsilon\eta\right)=0, \end{align}
it is then sufficient to show that
\small
\begin{align}\nonumber
&\lim_{\delta\rightarrow0}\limsup_NP\!\!\left(\!\!\sup_{\phi,\phi'\in\mathcal{R}_l\atop|\phi-\phi'|<\delta}\!\!\!\frac{1}{\sqrt{N}}\left|{\bf g}^H(\phi)\hat{\bf C}_N^{-1}{\bf z}_0\!-\!{\bf g}^H(\phi')\hat{\bf C}_N^{-1}{\bf z}_0\right|>\varepsilon'\!\!\right)\\ \label{eq:conv_v}
&=0
\end{align}
\normalsize
for any $\varepsilon'>0$ and similarly for \small${\bf g}^H(\phi')\hat{\bf C}_N^{-1}{\bf g}(\phi')-{\bf g}^H(\phi)\hat{\bf C}_N^{-1}{\bf g}(\phi)$\normalsize. Let us prove (\ref{eq:conv_v}), the other result following essentially the same line of arguments. For this, by Kallenberg \cite[Corollary 16.9]{kallenberg1997foundations}, it is sufficient to prove, say
\begin{align}\label{eq:expect}
\sup_{\phi,\phi'\in\mathcal{R}_l\atop|\phi\neq\phi'|}\sup_N\frac{E\left[\frac{1}{N}|{\bf g}^H(\phi)\hat{\bf C}_N^{-1}{\bf z}_0-{\bf g}^H(\phi')\hat{\bf C}_N^{-1}{\bf z}_0|^2\right]}{|\phi-\phi'|^2}<\infty.
\end{align}
Since
\begin{align}\nonumber
&\frac{1}{N}E\left[\left|{\bf g}^H(\phi)\hat{\bf C}_N^{-1}{\bf z}_0-{\bf g}^H(\phi')\hat{\bf C}_N^{-1}{\bf z}_0\right|^2\right]\\ \nonumber
&=\frac{1}{N}\|{\bf g}(\phi)-{\bf g}(\phi')\|^2E\left[\left\|\hat{\bf C}_N^{-2}\right\|\left\|{\bf z}_0{\bf z}_0^H\right\|\right],
\end{align}
and
$E\left[\left\|\hat{\bf C}_N^{-2}\right\|\left\|{\bf z}_0{\bf z}_0^H\right\|\right]<\infty$,
to prove (\ref{eq:expect}), we only need to prove
\begin{align}\label{eq:phi}
\frac{\frac{1}{N}\|{\bf g}(\phi)-{\bf g}(\phi')\|^2}{|\phi-\phi'|^2}<\infty.
\end{align}
Since
\begin{align}\nonumber
\frac{1}{N}\|{\bf g}(\phi)-{\bf g}(\phi')\|^2=\frac{1}{N}\sum_{m=1}^N|g_m(\phi)-g_m(\phi')|^2,
\end{align}
we first focus on analyzing $|g_m(\phi)-g_m(\phi')|$ for $i=1,\ldots,N$.
Denote $\phi'=\phi+\tau$,
\begin{align}\nonumber
&|g_m(\phi)-g_m(\phi')|=\left|-\frac{1}{2}\cosh(2\mu_m\phi-\mu_mx_m)\right.\\ \nonumber
&\left.\qquad\qquad\qquad\qquad\quad+\frac{1}{2}\cosh(2\mu_m\phi+2\mu_m\tau-\mu_mx_m)\right|\\ \nonumber
&\!\!=\left|\frac{1}{2}\cosh(2\mu_m\phi-\mu_mx_m)(\cosh2\mu_mm\tau-1)\right.\\\nonumber
&\left.\qquad+\frac{1}{2}\sinh(2\mu_m\phi-\mu_mx_m)\sinh2\mu_m\tau\right| \\ \label{eq:neq}
&\!\!\leq\!\frac{1}{2}\left|\cosh\mu_mL(\cosh2\mu_m\tau-1)\!+\!\sinh\mu_mL\sinh2\mu\tau\right|,
\end{align}
where the equality in (\ref{eq:neq}) is obtained when $\phi=L$ and $x_m=L$.

Therefore we establish the following inequality
\begin{align}\nonumber
\frac{\frac{1}{N}\sum_{m=1}^N|g_m(\phi)-g_m(\phi')|^2}{|\phi-\phi'|^2}\leq\Delta
\end{align}
where
\small
\begin{align}\nonumber
\Delta = \frac{\frac{1}{N}\sum_{m=1}^N\frac{1}{4}|\cosh\mu_mL(\cosh2\mu_m\tau-1)+\sinh\mu_mL\sinh2\mu_m\tau|^2}{\tau^2}.
\end{align}
\normalsize

The Tyler expansions of $\cosh 2\mu_m\tau$ and $\sinh 2\mu_m\tau$ are
\begin{align}\nonumber
\cosh 2\mu_m\tau = & 1+\frac{(2\mu_m\tau)^2}{2!}+\frac{(2\mu_m\tau)^4}{4!}+\frac{(2\mu_m\tau)^6}{6!} \\ \nonumber
&+\frac{(2\mu_m\tau)^8}{8!}+\cdots, \\ \nonumber
\sinh 2\mu_m\tau = & 2\mu_m\tau+\frac{(2\mu_m\tau)^3}{3!}+\frac{(2\mu_m\tau)^5}{5!} \\ \nonumber
&+\frac{(2\mu_m\tau)^7}{7!}+\frac{(2\mu_m\tau)^9}{9!}+\cdots.
\end{align}
By plugging in these Tyler expansions in $\Delta$, we obtain
\small
\begin{align}\nonumber
\Delta=\frac{1}{N}\sum_{m=1}^N\frac{1}{4}\left|\cosh\mu_mL\left(\frac{(2\mu_m)^2\tau}{2!}+\frac{(2\mu_m)^4\tau^3}{4!}+\frac{(2\mu_m)^6\tau^5}{6!}\right.\right.\\\nonumber
\left.\left.+\frac{(2\mu_m)^8\tau^7}{8!}+\cdots\right)+\sinh\mu_mL\left(2\mu_m+\frac{(2\mu_m)^3\tau^2}{3!}\right.\right.\\\nonumber
\left.\left.+\frac{(2\mu_m)^5\tau^4}{5!}+\frac{(2\mu_m)^7\tau^6}{7!}+\frac{(2\mu_m)^9\tau^8}{9!}+\cdots.\right)\right|^2.
\end{align}
\normalsize
It can be observed that $\Delta$ is an increasing function of $\tau$. Since $\tau\leq L$, we have
\small
\begin{align}\nonumber
\{\Delta(\tau)\}_{\rm max}=\Delta(L)=\frac{1}{L^2}\frac{1}{N}\sum_{m=1}^N\frac{1}{4}\left|\cosh\mu_mL(\cosh2\mu_mL-1)\right.\\\nonumber
\left.+\sinh\mu_mL\sinh2\mu_mL\right|^2<\infty.
\end{align}
\normalsize
Therefore we have proven (\ref{eq:phi}) and (\ref{eq:expect}), and also complete the proof of (\ref{eq:lipschitz}).

Getting back to our original problem, let us now take $\varepsilon>0$ arbitrary, $\phi_1<\ldots<\phi_J$ be a regular sampling of $\mathcal{R}_l$, and $\delta=\frac{L}{J}$. Then by (\ref{eq:result1}), J being fixed, for all $n>n_0(\varepsilon)$,
\begin{align}\label{eq:max}
\max_{1\leq j\leq J}\left|P(L(\phi_j)>\alpha)-Q_1\left(\frac{\alpha}{\sigma^2(\phi_j)}\right)\right|<\varepsilon.
\end{align}
Also, from (\ref{eq:lipschitz}), for small enough $\delta$,
\begin{align}\nonumber
\max_{1\leq j\leq J}P\left(\sup_{\phi\in\mathcal{R}_l\atop|\phi-\phi_j|<\delta}\left|L(\phi)-L(\phi_j)\right|>\alpha\xi\right)\\\nonumber
\leq P\left(\sup_{\phi,\phi'\in\mathcal{R}_l\atop|\phi-\phi'|<\delta}\left|L(\phi)-L(\phi')\right|>\alpha\xi\right)
<\varepsilon
\end{align}
for all large $n>n_0'(\varepsilon,\xi)>n_0(\varepsilon)$ where $\xi>0$ is also taken arbitrarily small. Thus we have, for each $\phi\in\mathcal{R}_l$ and for $n>n_0'(\varepsilon,\xi)$,
\small
\begin{align}\nonumber
P(L(\phi)>\alpha)&\leq P\left(L(\phi_i)>\alpha(1-\xi)\right)\!+\!P\left(\left|L(\phi)-L(\phi')\right|>\alpha\xi\right) \\\nonumber
&\leq P(L(\phi_i)>\alpha(1-\xi))+\varepsilon
\end{align}
\normalsize
for $i\leq J$ the unique index such that $|\phi-\phi_i|<\delta$ and where the inequality holds uniformly on $\phi\in\mathcal{R}_l$.

Similarly, reversing the roles of $\phi$ and $\phi'$,
\begin{align}\nonumber
P(L(\phi)>\alpha)\geq P(L(\phi_i)>\alpha(1+\xi))-\varepsilon.
\end{align}
As a consequence, by (\ref{eq:max}), for $n>n_0'(\varepsilon,\xi)$, uniformly on $\phi\in\mathcal{R}_l$,
\begin{align}\nonumber
P(L(\phi)>\alpha)&\leq Q_1\left(\frac{\alpha(1-\xi)}{2\sigma^2(\phi_i)}\right)+2\varepsilon \\\nonumber
P(L(\phi)>\alpha)&\geq Q_1\left(\frac{\alpha(1+\xi)}{2\sigma^2(\phi_i)}\right)-2\varepsilon
\end{align}
which, by continuity of $Q_1$ and $\phi\mapsto\sigma^2$, letting $\xi$ and $\delta$ small enough (up to growing $n_0'(\varepsilon,\xi)$), leads to
\begin{align}\nonumber
\sup_{\phi\in\mathcal{R}_l}\left|P(L(\phi)>\alpha)-Q_1\left(\frac{\alpha}{2\sigma^2(\phi)}\right)\right|\leq3\varepsilon
\end{align}
for all $n_0'(\varepsilon,\xi)$, which completes the uniform convergence across $\phi\in\mathcal{R}_l$.

\subsection{Proof of Theorem \ref{th:Pd}}\label{appx:th_Pd}
We first study the asymptotic behavior of the detection probability for fixed $\rho\in\mathcal{R}_{\kappa}$ and $\phi\in\mathcal{R}_l$. As shown in \cite{kammoun2018optimal}, under $H_1$, $\frac{1}{\sqrt{N}}{\rm Re}({\bf g}^H(\phi)\hat{\bf C}_N^{-1}(\rho){\bf z}_0)$ behaves asymptotically as a Gaussian variable with mean $\mu=\frac{s}{\sqrt{N}\rho}{\bf g}^H(\phi){\bf Q}_N(\underline{\rho}){\bf g}(\phi)$ and variance $\nu^2\!=\!\frac{1}{2\rho^2N}\frac{{\bf g}^H(\phi){\bf C}_N{\bf Q}_N^2(\underline{\rho}){\bf g}(\phi)}{1-cm_N(-\underline{\rho})^2(1-\underline{\rho})^2\frac{1}{N}{\rm tr}{\bf C}_N^2{\bf Q}_N^2(\underline{\rho})}$ as $N,K\!\!\rightarrow\infty$, with $c_N\!=\!N/K\!\rightarrow\!c\in(0,1)$. Thus, $\frac{1}{N}{\rm Re}^2\{{\bf g}^H(\phi)\hat{\bf C}_N^{-1}(\rho){\bf z}_0\}$ behaves asymptotically as a noncentral chi-squared random variable with degree of freedom $1$, parameterized by the location $\mu^2$ and scale $\nu$. Combining this result and (\ref{eq:conv_gphi}) along with Slutsky's lemma, we conclude that, under $H_1$, $L(\rho,\phi)$ in (\ref{eq:L_rho}) is also asymptotically equivalent to a noncentral chi-squared random variable with degree of freedom $1$, but with location $\frac{\mu^2}{{\bf g}^H(\phi){\bf Q}_N(\underline{\rho}){\bf g}(\phi)}$ and scale $\sigma$.

Defining $\beta(\rho,\phi)=\frac{\mu}{\sigma\sqrt{{\bf g}^H(\phi){\bf Q}_N(\underline{\rho}){\bf g}(\phi)}}$\normalsize, we therefore conclude, for fixed $\rho\in\mathcal{R}_{\kappa}$ and $\phi\in\mathcal{R}_l$, that
\begin{align}\nonumber
\left|\mathbb{P}\left[L(\rho,\phi)>\alpha|H_1\right]-Q_2\left(\beta^2(\rho,\phi)), \frac{\alpha}{\sigma^2(\rho)}\right)\right|\rightarrow0.
\end{align}
As before, the generalization to include uniform convergence across $\rho\in\mathcal{R}_{\kappa}$ and $\phi\in\mathcal{R}_l$ can be derived by following the same procedure as in \cite{couillet2016second} and the proof of Theorem \ref{th:Pfa}, and is therefore again not reproduced.

\subsection{Proof of Proposition \ref{th:sigma}}\label{appx:th_sigma}
The proof consists of two steps. Firstly we prove that for a fixed $\phi\in\mathcal{R}_l$, the following convergence result holds:
\begin{align}\label{eq:convrho}
\sup_{\rho\in\mathcal{R}_{\kappa}}\left|\hat{\sigma}^2(\rho,\phi)-\sigma^2(\rho,\phi)\right|\stackrel{a.s.}\longrightarrow0.
\end{align}
Then the uniform convergence over $\phi\in\mathcal{R}_l$ is deducted, which completes the proof.

In the first step, we start by showing that $\hat{\sigma}^2(1,\phi)$ is well defined. It is easy to observe that $\hat{\sigma}^2(\rho,\phi)$ in (\ref{eq:sigma_est}) is undefined (zero over zero) when $\rho=1$. We use l'Hopital's rule to obtain the value of $\hat{\sigma}^2(\rho,\phi)$ when $\rho$ approaches $1$. Define $\hat{\sigma}^2(\rho,\phi)=\frac{h(\rho,\phi)}{w(\rho)}$ with $h(\rho,\phi)$ and $w(\rho)$ given by
\begin{align}\nonumber
h(\rho,\phi)=1-\frac{\rho{\bf g}^H(\phi)\hat{\bf C}_N^{-2}(\rho){\bf g}(\phi)}{{\bf g}^H(\phi)\hat{\bf C}_N^{-1}(\rho){\bf g}(\phi)}
\end{align}
and
\begin{align}\nonumber
w(\rho)=\frac{2(1-\rho)N}{{\rm tr}({\bf R}_N)}\left(1-c_N+c_N\rho\frac{1}{N}{\rm tr}\hat{\bf C}_N^{-1}(\rho)\right)^2.
\end{align}
By a uniform variation of l'Hopital's rule \cite[Lemma 13]{kammoun2018optimal}, we have
\begin{align}\nonumber
\lim_{\rho\uparrow1}\limsup_N\left|\hat{\sigma}^2(\rho,\phi)-\frac{h'(1,\phi)}{w'(1)}\right|\stackrel{\rm a.s.}\longrightarrow0.
\end{align}
Using the differentiation rules $\frac{d}{d\rho}\hat{\bf C}_N^{-1}(\rho)\!=\!-\hat{\bf C}_N^{-2}(\rho)(-{\bf R}_N+{\bf I}_N)$ and $\frac{d}{d\rho}\hat{\bf C}_N^{-2}(\rho)=-\hat{\bf C}_N^{-3}(\rho)(-{\bf R}_N+{\bf I}_N)$ \cite{kammoun2018optimal}, we then prove
\begin{align}\nonumber
\lim_{\rho\uparrow1}\limsup_N\left|\hat{\sigma}^2(\rho,\phi)-\frac{{\bf g}^H(\phi){\bf R}_N{\bf g}(\phi)}{2{\bf g}^H(\phi){\bf g}(\phi)}\right|\stackrel{\rm a.s.}\longrightarrow0.
\end{align}
Now, using the fact that as $N, K\rightarrow\infty$, with $c_N\rightarrow c\in(0,1)$, $\frac{1}{N}{\bf g}^H(\phi){\bf R}_N{\bf g}(\phi)-\frac{1}{N}{\bf g}^H(\phi){\bf C}_N{\bf g}(\phi)\stackrel{\rm a.s.}\longrightarrow0$ \cite{kammoun2018optimal}, we obtain
\begin{align}\nonumber
\lim_{\rho\uparrow1}\limsup_N\left|\hat{\sigma}^2(\rho,\phi)-\frac{{\bf g}^H(\phi){\bf C}_N{\bf g}(\phi)}{2{\bf g}^H(\phi){\bf g}(\phi)}\right|\stackrel{\rm a.s.}\longrightarrow0 \; .
\end{align}
 Since $\sigma^2(1,\phi)=\frac{{\bf g}^H(\phi){\bf C}_N{\bf g}(\phi)}{2{\bf g}^H(\phi){\bf g}(\phi)}$, we have thus proved as $N, K\rightarrow\infty$, with $c_N\rightarrow c\in(0,1)$, $\left|\hat{\sigma}^2(1,\phi)-\sigma^2(1,\phi)\right|\stackrel{\rm a.s.}\longrightarrow0$ where $\hat{\sigma}^2(1,\phi)=\lim_{\rho\uparrow1}\hat{\sigma}^2(\rho,\phi)$.

It then suffices to prove (\ref{eq:convrho}) when $\rho$ belongs to the set $\tilde{\mathcal{R}}_{\kappa}\triangleq[\kappa,1-\kappa]$.
By (\ref{eq:conv_gphi}), we could obtain the consistent estimator of the first part of $\sigma^2(\rho,\phi)$, that is $\frac{1}{2\rho}\frac{1}{{\bf g}^H(\phi){\bf Q}_N(\underline{\rho}){\bf g}(\phi)}$. For the remaining part of $\sigma^2(\rho,\phi)$, $\frac{{\bf g}^H(\phi){\bf C}_N{\bf Q}_N^2(\underline{\rho}){\bf g}(\phi)}{1-cm_N^2(-\underline{\rho})(1-\underline{\rho})^2\frac{1}{N}{\rm tr}{\bf C}_N^2{\bf Q}_N^2(\underline{\rho})}$,
the following convergence results in \cite{couillet2016second} are exploited:
\small
\begin{align}\nonumber
\sup_{\rho\in\tilde{\mathcal{R}}_{\kappa}}&\!\left|\!\frac{1}{N}\frac{{\bf g}^H(\phi)\hat{\bf C}_N^{-1}(\rho){\bf g}(\phi)\!-\!\rho{\bf g}^H(\phi)\hat{\bf C}_N^{-2}(\rho){\bf g}(\phi)}{(1-\underline{\rho})m_N^2(-\underline{\rho})}\!\left(\!\rho\!+\!\frac{(1-\rho)N}{{\rm tr}({\bf R}_N)}\!\right)\right.\\ \label{eq:3}
&~\left.-\frac{1}{N}\frac{{\bf g}^H(\phi){\bf C}_N{\bf Q}_N^2(\underline{\rho}){\bf g}(\phi)}{1-cm_N^2(-\underline{\rho})(1-\underline{\rho})^2\frac{1}{N}{\rm tr}{\bf C}_N^2{\bf Q}_N^2(\underline{\rho})}\right|\stackrel{\rm a.s.}\longrightarrow0
\end{align}
\normalsize
and
\small
\begin{align}\nonumber
\sup_{\rho\in\tilde{\mathcal{R}}_{\kappa}}&\left|\left(\frac{1-c_N}{\underline{\rho}}+c_N\frac{1}{N}{\rm tr}\hat{\bf C}_N^{-1}(\rho)\left(\rho+\frac{(1-\rho)N}{{\rm tr}({\bf R}_N)}\right)\right) \right.\\\label{eq:m_rho}
&~\left.-m_N(-\underline{\rho})\right|\stackrel{\rm a.s.}\longrightarrow0.
\end{align}
\normalsize
By combining (\ref{eq:3}) and (\ref{eq:m_rho}), we have
\small
\begin{align}\nonumber
&\sup_{\rho\in\tilde{\mathcal{R}}_{\kappa}}\left|\frac{1}{N}\frac{{\bf g}^H(\phi){\bf C}_N{\bf Q}_N^2(\underline{\rho}){\bf g}(\phi)}{1-cm_N^2(-\underline{\rho})(1-\underline{\rho})^2\frac{1}{N}{\rm tr}{\bf C}_N^2{\bf Q}_N^2(\underline{\rho})}\right.\\ \nonumber
&\left.-\frac{1}{N}\frac{{\rm tr}({\bf R}_N)}{(1-\rho)N}\frac{{\bf g}^H(\phi)\hat{\bf C}_N^{-1}(\rho){\bf g}(\phi)-\rho{\bf g}^H(\phi)\hat{\bf C}_N^{-2}(\rho){\bf g}(\phi)}{\left(1-c_N+c_N\rho\frac{1}{N}{\rm tr}\hat{\bf C}_N^{-1}(\rho)\right)^2}\right|\stackrel{\rm a.s.}\longrightarrow0.
\end{align}
\normalsize
Together with (\ref{eq:conv_gphi}), we prove the uniform convergence (\ref{eq:convrho}) over $\rho\in\mathcal{R}_{\kappa}$.

In the second step, we prove the uniform convergence over $\phi\in\mathcal{R}_l$. To simplify notations, we again drop the parameter $\rho$, that is, we aim to prove, as $N, K\rightarrow\infty$, with $c_N\rightarrow c\in(0,1)$,
\begin{align}\label{eq:uniformsigma}
\sup_{\phi\in\mathcal{R}_l}\left|\hat{\sigma}^2(\phi)-\sigma^2(\phi)\right|\stackrel{a.s.}\longrightarrow0.
\end{align}
From the definition of uniform convergence, this amounts to showing that for some $C>0$ and any given $\varepsilon>0$,
\begin{align}\label{eq:C}
\sup_{\phi\in\mathcal{R}_l}\left|\hat{\sigma}^2(\phi)-\sigma^2(\phi)\right|<C\varepsilon
\end{align}
for all large $K$ almost surely.

Taking $\phi_1<\ldots<\phi_J$ be a regular sampling of $\mathcal{R}_l$, and $\delta=\frac{L}{J}$, there exist $\phi_i$ that satisfies $|\phi-\phi_i|<\delta$. With this, we can write:
\begin{align}\nonumber
&\sup_{\phi\in\mathcal{R}_l}|\hat{\sigma}^2(\phi)-\sigma^2(\phi)|
\leq\sup_{\phi\in\mathcal{R}_l}\left\{|\hat{\sigma}^2(\phi)-\hat{\sigma}^2(\phi_i)| \right.\\\nonumber
&\left.\qquad\qquad\qquad+|\sigma^2(\phi_i)-\sigma^2(\phi)|+|\hat{\sigma}^2(\phi_i)-\sigma^2(\phi_i)|\right\} \\ \nonumber
&\leq
\sup_{\phi\in\mathcal{R}_l}|\sigma^2(\phi_i)-\sigma^2(\phi)|+\sup_{\phi\in\mathcal{R}_l}|\hat{\sigma}^2(\phi)-\hat{\sigma}^2(\phi_i)|\\\label{eq:3terms}
&\qquad+\max_i|\hat{\sigma}^2(\phi_i)-\sigma^2(\phi_i)|.
\end{align}
Hence, it follows that the relation (\ref{eq:C}) would be established upon proving that, for certain $C_1>0$, $C_2>0$ and $C_3>0$, we have $\sup_{\phi\in\mathcal{R}_l}|\sigma^2(\phi_i)-\sigma^2(\phi)|<C_2\varepsilon$, $\sup_{\phi\in\mathcal{R}_l}|\hat{\sigma}^2(\phi)-\hat{\sigma}^2(\phi_i)|<C_1\varepsilon$, $\max_i|\hat{\sigma}^2(\phi_i)-\sigma^2(\phi_i)|<C_3\varepsilon$ for all large $K$ almost surely.

To establish the first bound, we start by using (\ref{eq:sigma}) to write
\small
\begin{align}\nonumber
&|\sigma^2(\phi_i)-\sigma^2(\phi)|=q*\left|\frac{{\bf g}^H(\phi_i){\bf C}_N{\bf Q}_N^2{\bf g}(\phi_i)}{{\bf g}^H(\phi_i){\bf Q}_N{\bf g}(\phi_i)}-\frac{{\bf g}^H(\phi){\bf C}_N{\bf Q}_N^2{\bf g}(\phi)}{{\bf g}^H(\phi){\bf Q}_N{\bf g}(\phi)}\right| \\ \nonumber
&\!\!\!=\!\!\frac{q}{{\bf g}^H(\phi_i){\bf Q}_N{\bf g}(\phi_i){\bf g}^H(\phi){\bf Q}_N{\bf g}(\phi)}\!\!\left|{\bf g}^H(\phi_i){\bf C}_N{\bf Q}_N^2{\bf g}(\phi_i){\bf g}^H(\phi){\bf Q}_N{\bf g}(\phi)\right.\\ \nonumber
&\left.\quad-{\bf g}^H(\phi){\bf C}_N{\bf Q}_N^2{\bf g}(\phi){\bf g}^H(\phi_i){\bf Q}_N{\bf g}(\phi_i)\right|,
\end{align}
\normalsize
where
$q = \frac{1}{2\rho}\frac{1}{1-cm_N^2(-\underline{\rho})(1-\underline{\rho})^2\frac{1}{N}{\rm tr}{\bf C}_N^2{\bf Q}_N^2(\underline{\rho})}$.

Rewrite $|\sigma^2(\phi_i)-\sigma^2(\phi)|=q*\frac{A}{B}$, where
\small
\begin{align}\nonumber
A&\!\triangleq\!\!\frac{1}{N^2}{\bf g}^H\!(\phi){\bf Q}_N{\bf g}(\phi)[{\bf g}^H\!(\phi_i){\bf C}_N{\bf Q}_N^2{\bf g}(\phi_i)\!-\!{\bf g}^H\!(\phi){\bf C}_N{\bf Q}_N^2{\bf g}(\phi)]\\\nonumber
&~+\frac{1}{N^2}[{\bf g}^H\!(\phi){\bf Q}_N{\bf g}(\phi)\!-\!{\bf g}^H\!(\phi_i){\bf Q}_N{\bf g}(\phi_i)]{\bf g}^H\!(\phi){\bf C}_N{\bf Q}_N^2{\bf g}(\phi), \\\nonumber
B&\triangleq\frac{1}{N^2}{\bf g}^H(\phi_i){\bf Q}_N{\bf g}(\phi_i){\bf g}^H(\phi){\bf Q}_N{\bf g}(\phi).
\end{align}
\normalsize
We first deal with $A$. Since
\begin{align}\nonumber
&\frac{1}{N}[{\bf g}^H(\phi_i){\bf C}_N{\bf Q}_N^2{\bf g}(\phi_i)-{\bf g}^H(\phi){\bf C}_N{\bf Q}_N^2{\bf g}(\phi)]\\ \nonumber
&=\frac{1}{N}({\bf g}(\phi_i)-{\bf g}(\phi))^H{\bf C}_N{\bf Q}_N^2({\bf g}(\phi)+{\bf g}(\phi_i)) \\\nonumber
&\leq\frac{1}{\sqrt{N}}\|{\bf g}(\phi)-{\bf g}(\phi_i)\|\|{\bf C}_N{\bf Q}_N^2\|\frac{1}{\sqrt{N}}\|{\bf g}(\phi)+{\bf g}(\phi_i)\|
\end{align}
and
\begin{align}\nonumber
&\frac{1}{N}[{\bf g}^H(\phi){\bf Q}_N{\bf g}(\phi)-{\bf g}^H(\phi_i){\bf Q}_N{\bf g}(\phi_i)] \\\nonumber
&=\frac{1}{N}({\bf g}(\phi)-{\bf g}(\phi_i))^H{\bf Q}_N({\bf g}(\phi)+{\bf g}(\phi_i)) \\\nonumber
&\leq\frac{1}{\sqrt{N}}\|{\bf g}(\phi)-{\bf g}(\phi_i)\|\|{\bf Q}_N\|\frac{1}{\sqrt{N}}\|{\bf g}(\phi)+{\bf g}(\phi_i)\|,
\end{align}
we have
\begin{align}\nonumber
A\leq&\frac{1}{\sqrt{N}}\|{\bf g}(\phi)-{\bf g}(\phi_i)\| \\\nonumber
&\times\left(\frac{1}{N}{\bf g}^H(\phi){\bf Q}_N{\bf g}(\phi)\|{\bf C}_N{\bf Q}_N^2\|\frac{1}{\sqrt{N}}\|{\bf g}(\phi)+{\bf g}(\phi_i)\| \right.\\\nonumber
&\left.+\|{\bf Q}_N\|\frac{1}{\sqrt{N}}\|{\bf g}(\phi)+{\bf g}(\phi_i)\|\frac{1}{N}{\bf g}^H(\phi){\bf C}_N{\bf Q}_N^2{\bf g}(\phi)\right).
\end{align}

Denote $\phi_i=\phi+\tau$, $|\tau|<\delta<L$, we obtain
\begin{align}\nonumber
&g_m(\phi_i)-g_m(\phi)=\cosh(2\mu_m\phi-\mu_mx_m)\sinh^2\mu_m\tau \\\nonumber
&\qquad\qquad+\sinh(2\mu_m\phi-\mu_mx_m)\sinh\mu_m\tau\cosh\mu_m\tau \\\nonumber
&=\sinh\mu_m\tau[\cosh(2\mu_m\phi-\mu_mx_m)\sinh\mu_m\tau \\\nonumber
&\qquad\qquad\qquad+\sinh(2\mu_m\phi-\mu_mx_m)\cosh\mu_m\tau]\\\nonumber
&<\sinh\mu_m\tau[\cosh(2\mu_m\phi-\mu_mx_m)\sinh\mu_mL\\\nonumber
&\qquad\qquad\qquad+\sinh(2\mu_m\phi-\mu_mx_m)\cosh\mu_mL].
\end{align}
By taking $\delta$ satisfies $\max_m\sinh\mu_m\delta<\varepsilon$, we obtain, for each $m$, $m=1,\ldots, N$,
\begin{align}\nonumber
g_m(\phi_i)-g_m(\phi)<h_m\varepsilon,
\end{align}
where $h_m = \cosh(2\mu_m\phi-\mu_mx_m)\sinh\mu_mL+\sinh(2\mu_m\phi-\mu_mx_m)\cosh\mu_mL$.

Since
\begin{align}\nonumber
\frac{1}{\sqrt{N}}\|{\bf g}(\phi)-{\bf g}(\phi_i)\|&=\frac{1}{\sqrt{N}}\sqrt{\sum_{m=1}^N|g_m(\phi)-g_m(\phi_i)|^2} \\ \nonumber
&<\frac{1}{\sqrt{N}}\sqrt{\sum_{m=1}^Nh_m^2}\varepsilon,
\end{align}
we obtain
\small
\begin{align}\nonumber
&A<\left(\frac{1}{N}{\bf g}^H(\phi){\bf Q}_N{\bf g}(\phi)\|{\bf C}_N{\bf Q}_N^2\|\frac{1}{\sqrt{N}}\|{\bf g}(\phi)+{\bf g}(\phi_i)\| \right.\\\nonumber
&\left.\!\!\!+\|{\bf Q}_N\|\frac{1}{\sqrt{N}}\|{\bf g}(\phi)+{\bf g}(\phi_i)\|\frac{1}{N}{\bf g}^H(\phi){\bf C}_N{\bf Q}_N^2{\bf g}(\phi)\!\!\right)\!\!\frac{1}{\sqrt{N}}\sqrt{\sum_{m=1}^Nh_m^2}\varepsilon \\\nonumber
&<\frac{2}{N}\|{\bf g}(\phi)\|^2\frac{1}{\sqrt{N}}\|{\bf g}(\phi)+{\bf g}(\phi_i)\|\|{\bf Q}_N\|\|{\bf C}_N{\bf Q}_N^2\|\frac{1}{\sqrt{N}}\sqrt{\sum_{m=1}^Nh_m^2}\varepsilon.
\end{align}
\normalsize
As $\frac{1}{\sqrt{N}}\|{\bf g}(\phi)\|$, $\|{\bf C}_N\|$ and $\|{\bf Q}_N\|$ are bounded, we have $A< p_1\varepsilon$ for some constant $p_1$.

Similarly, since
\begin{align}\nonumber
B\geq\frac{1}{N^2}\|{\bf g}(\phi_i)\|^2\|{\bf g}(\phi)\|^2\|{\bf Q}_N\|^2
\end{align}
and $\frac{1}{\sqrt{N}}\|{\bf g}(\phi)\|$ and $\|{\bf Q}_N\|$ is bounded, we have $B>p_2$, for some constant $p_2$.
Therefore, we have established the desired property
\begin{align}\label{eq:term1}
|\sigma^2(\phi_i)-\sigma^2(\phi)|=q*\frac{A}{B}<\frac{qp_1}{p_2}\varepsilon.
\end{align}

We now turn to deriving the analogous result for the second term in (\ref{eq:3terms}). To this end, similar to before, we start with
\small
\begin{align}\nonumber
&|\hat{\sigma}^2(\phi)-\hat{\sigma}^2(\phi_i)|=r\times\left|\frac{{\bf g}^H(\phi)\hat{\bf C}_N^{-2}{\bf g}(\phi)}{{\bf g}^H(\phi)\hat{\bf C}_N^{-2}{\bf g}(\phi)}-\frac{{\bf g}^H(\phi_i)\hat{\bf C}_N^{-2}{\bf g}(\phi_i)}{{\bf g}^H(\phi_i)\hat{\bf C}_N^{-2}{\bf g}(\phi_i)}\right| \\\nonumber
&=\frac{r}{\frac{1}{N^2}{\bf g}^H(\phi)\hat{\bf C}_N^{-1}{\bf g}(\phi){\bf g}^H(\phi_i)\hat{\bf C}_N^{-1}{\bf g}(\phi_i)}\\\nonumber
&\qquad\times\left|\frac{1}{N^2}[{\bf g}^H(\phi)\hat{\bf C}_N^{-2}{\bf g}(\phi){\bf g}^H(\phi_i)\hat{\bf C}_N^{-1}{\bf g}(\phi_i)- \right.\\ \nonumber
&\qquad\qquad\left.{\bf g}^H(\phi_i)\hat{\bf C}_N^{-2}{\bf g}(\phi_i){\bf g}^H(\phi)\hat{\bf C}_N^{-1}{\bf g}(\phi)]\right|,
\end{align}
\normalsize
where $r=\frac{{\rm tr}({\bf R}_N)}{2(1-\rho)N}\frac{1}{{\left(1-c_N+c_N\rho\frac{1}{N}{\rm tr}\hat{\bf C}_N^{-1}(\rho)\right)^2}}$.

Rewrite $|\hat{\sigma}^2(\phi)-\hat{\sigma}^2(\phi_i)|=r*\frac{D}{E}$, where
\begin{align}\nonumber
D&\triangleq\frac{1}{N^2}{\bf g}^H(\phi)\hat{\bf C}_N^{-2}{\bf g}(\phi)[{\bf g}^H(\phi_i)\hat{\bf C}_N^{-1}{\bf g}(\phi_i)-{\bf g}^H(\phi)\hat{\bf C}_N^{-1}{\bf g}(\phi)]\\\nonumber
&+\frac{1}{N^2}[{\bf g}^H(\phi)\hat{\bf C}_N^{-2}{\bf g}(\phi)-{\bf g}^H(\phi_i)\hat{\bf C}_N^{-2}{\bf g}(\phi_i)]{\bf g}^H(\phi)\hat{\bf C}_N^{-1}{\bf g}(\phi), \\\nonumber
E&\triangleq\frac{1}{N^2}{\bf g}^H(\phi)\hat{\bf C}_N^{-1}{\bf g}(\phi){\bf g}^H(\phi_i)\hat{\bf C}_N^{-1}{\bf g}(\phi_i).
\end{align}
We first deal with $D$. Since
\begin{align}\nonumber
&\frac{1}{N}[{\bf g}^H(\phi_i)\hat{\bf C}_N^{-1}{\bf g}(\phi_i)-{\bf g}^H(\phi)\hat{\bf C}_N^{-1}{\bf g}(\phi)]\\\nonumber
&=\frac{1}{N}({\bf g}(\phi_i)-{\bf g}(\phi))^H\hat{\bf C}_N^{-1}({\bf g}(\phi)+{\bf g}(\phi_i)) \\\nonumber
&\leq\frac{1}{\sqrt{N}}\|{\bf g}(\phi)-{\bf g}(\phi_i)\|\|\hat{\bf C}_N^{-1}\|\frac{1}{\sqrt{N}}\|{\bf g}(\phi)+{\bf g}(\phi_i)\|
\end{align}
and
\begin{align}\nonumber
&\frac{1}{N}[{\bf g}^H(\phi_i)\hat{\bf C}_N^{-2}{\bf g}(\phi_i)-{\bf g}^H(\phi)\hat{\bf C}_N^{-2}{\bf g}(\phi)] \\\nonumber
&=\frac{1}{N}({\bf g}(\phi_i)-{\bf g}(\phi))^H\hat{\bf C}_N^{-2}({\bf g}(\phi)+{\bf g}(\phi_i)) \\\nonumber
&\leq\frac{1}{\sqrt{N}}\|{\bf g}(\phi)-{\bf g}(\phi_i)\|\|\hat{\bf C}_N^{-2}\|\frac{1}{\sqrt{N}}\|{\bf g}(\phi)+{\bf g}(\phi_i)\|,
\end{align}
we have
\small
\begin{align}\nonumber
D\leq&\frac{1}{\sqrt{N}}\|{\bf g}(\phi)-{\bf g}(\phi_i)\|\\ \nonumber
&\times\left(\frac{1}{N}{\bf g}^H(\phi)\hat{\bf C}_N^{-2}{\bf g}(\phi)\|\hat{\bf C}_N^{-1}\|\frac{1}{\sqrt{N}}\|{\bf g}(\phi)+{\bf g}(\phi_i)\| \right.\\\nonumber
&\left.+\|\hat{\bf C}_N^{-2}\|\frac{1}{\sqrt{N}}\|{\bf g}(\phi)+{\bf g}(\phi_i)\|\frac{1}{N}{\bf g}^H(\phi)\hat{\bf C}_N^{-1}{\bf g}(\phi)\right)\\\nonumber
\leq&\frac{1}{\sqrt{N}}\|{\bf g}(\phi)-{\bf g}(\phi_i)\|\frac{2}{N}\|{\bf g}(\phi)\|^2\|\hat{\bf C}_N^{-2}\|\|\hat{\bf C}_N^{-1}\|\\\nonumber
&\times\frac{1}{\sqrt{N}}\|{\bf g}(\phi)+{\bf g}(\phi_i)\|.
\end{align}
\normalsize
As we have proved that
\begin{align}\nonumber
\frac{1}{\sqrt{N}}\|{\bf g}(\phi)-{\bf g}(\phi_i)\|<\frac{1}{\sqrt{N}}\sqrt{\sum_{m=1}^Nh_m^2}\varepsilon,
\end{align}
and $\frac{1}{\sqrt{N}}\|{\bf g}(\phi)\|$, $\|\hat{\bf C}_N^{-1}\|$ are bounded, we have $D< p_3\varepsilon$ for some constant $p_3$.

Similarly, since
\begin{align}\nonumber
E\geq\frac{1}{N^2}\|{\bf g}(\phi_i)\|^2\|{\bf g}(\phi)\|^2\|\hat{\bf C}_N^{-1}\|^2
\end{align}
and $\frac{1}{\sqrt{N}}\|{\bf g}(\phi)\|$ and $\|\hat{\bf C}_N^{-1}\|$ are bounded, we have $E>p_4$, for some constant $p_4$.
Therefore, we have established the desired property
\begin{align}\label{eq:term2}
|\hat{\sigma}^2(\phi)-\hat{\sigma}^2(\phi_i)|=r*\frac{D}{E} <\frac{rp_3}{p_4}\varepsilon.
\end{align}

Finally, we turn to deriving the analogous result (for all large $K$ almost surely) for the third term in (\ref{eq:3terms}). Since, as already established, for each $\phi_i$, as $N, K\rightarrow\infty$, with $c_N\rightarrow c\in(0,1)$, $\left|\hat{\sigma}^2(\phi_i)-\sigma^2(\phi_i)\right|\stackrel{a.s.}\longrightarrow0$, we have that for each $\phi_i$, $\left|\hat{\sigma}^2(\phi_i)-\sigma^2(\phi_i)\right|<\varepsilon$ for all large $K$ almost surely. Thus,
\begin{align}\nonumber
\max_i|\hat{\sigma}^2(\phi_i)-\sigma^2(\phi_i)|<\sum_{i=1}^J|\hat{\sigma}^2(\phi_i)-\sigma^2(\phi_i)|<J\varepsilon.
\end{align}
This, combined with (\ref{eq:term1}) and (\ref{eq:term2}) completes the proof that (\ref{eq:C}) holds, hence establishing the desired uniform convergence (\ref{eq:uniformsigma}).

\subsection{Proof of Theorem \ref{th:phi_est}} \label{appx:th_phi}
The proof relies on the following convergence results, which will be derived subsequently: As $N, K\rightarrow\infty$, with $c_N=N/K\rightarrow c\in(0,1)$,
\begin{align}\label{eq:conv_dmn}
\max_{\phi\in\mathcal{R}_l}\!\left|\frac{1}{N}{\bf g}^H(\phi){\bf R}_N^{-1}{\bf g}(\phi)\!-\!\frac{1}{1-c}\frac{1}{N}{\bf g}^H(\phi){\bf C}_N^{-1}{\bf g}(\phi)\right|\!\!\stackrel{\rm a.s.}\longrightarrow\!0
\end{align}
and
\begin{align}\nonumber
\max_{\phi\in\mathcal{R}_l]}&\left|\frac{1}{N}{\rm Re}^2\{{\bf g}^H(\phi){\bf R}_N^{-1}{\bf z}_0\} \right.\\ \label{eq:conv_nmt}
&~\left.-\frac{1}{(1-c)^2}\frac{1}{N}{\rm Re}^2\{{\bf g}^H(\phi){\bf C}_N^{-1}{\bf z}_0\}\right|\stackrel{\rm a.s.}\longrightarrow0.
\end{align}
We then have
\small
\begin{align}\nonumber
\max_{\phi\in\mathcal{R}_l}\left|\frac{{\rm Re}^2\{{\bf g}^H(\phi){\bf R}_N^{-1}{\bf z}_0\}}{{\bf g}^H(\phi){\bf R}_N^{-1}{\bf g}(\phi)}\!-\!\frac{1}{1-c}\frac{{\rm Re}^2\{{\bf g}^H(\phi){\bf C}_N^{-1}{\bf z}_0\}}{{\bf g}^H(\phi){\bf C}_N^{-1}{\bf g}(\phi)}\right|\!\stackrel{\rm a.s.}\longrightarrow\!0.
\end{align}
\normalsize
Denote $\hat{\phi}\in\argmax_{\phi\in[p_{\rm U}, p_{\rm D}]}\frac{{\rm Re}^2\{{\bf g}^H(\phi){\bf C}_N^{-1}{\bf z}_0\}}{{\bf g}^H(\phi){\bf C}_N^{-1}{\bf g}(\phi)}$.  Together with $\hat{\phi}_{\{{\bf R}_N,{\bf z}_0\}}\in\argmax_{\phi\in[p_{\rm U}, p_{\rm D}]}\frac{{\rm Re}^2\{{\bf g}^H(\phi){\bf R}_N^{-1}{\bf z}_0\}}{{\bf g}^H(\phi){\bf R}_N^{-1}{\bf g}(\phi)}$, the following inequalities hold true:
\small
\begin{align}\label{ieq:1}
\frac{{\rm Re}^2\{{\bf g}^H(\hat{\phi}_{\{{\bf R}_N,{\bf z}_0\}}){\bf R}_N^{-1}{\bf z}_0\}}{{\bf g}^H(\hat{\phi}_{\{{\bf R}_N,{\bf z}_0\}}){\bf R}_N^{-1}{\bf g}(\hat{\phi}_{\{{\bf R}_N,{\bf z}_0\}})}\geq\frac{{\rm Re}^2\{{\bf g}^H(\hat{\phi}){\bf R}_N^{-1}{\bf z}_0\}}{{\bf g}^H(\hat{\phi}){\bf R}_N^{-1}{\bf g}(\hat{\phi})}
\end{align}
\normalsize
and
\small
\begin{align}\label{ieq:2}
\frac{{\rm Re}^2\{{\bf g}^H(\hat{\phi}){\bf C}_N^{-1}{\bf z}_0\}}{{\bf g}^H(\hat{\phi}){\bf C}_N^{-1}{\bf g}(\hat{\phi})}\geq\frac{{\rm Re}^2\{{\bf g}^H(\hat{\phi}_{\{{\bf R}_N,{\bf z}_0\}}){\bf C}_N^{-1}{\bf z}_0\}}{{\bf g}^H(\hat{\phi}_{\rm SCM}){\bf C}_N^{-1}{\bf g}(\hat{\phi}_{\{{\bf R}_N,{\bf z}_0\}})} \; .
\end{align}
\normalsize
We also have
\small
\begin{align}\nonumber
&\left|\frac{{\rm Re}^2\{{\bf g}^H(\hat{\phi}_{\{{\bf R}_N,{\bf z}_0\}}){\bf R}_N^{-1}{\bf z}_0\}}{{\bf g}^H(\hat{\phi}_{\{{\bf R}_N,{\bf z}_0\}}){\bf R}_N^{-1}{\bf g}(\hat{\phi}_{\{{\bf R}_N,{\bf z}_0\}})}\right.\\\nonumber
&~\left.-\frac{1}{1-c}\frac{{\rm Re}^2\{{\bf g}^H(\hat{\phi}_{\{{\bf R}_N,{\bf z}_0\}}){\bf C}_N^{-1}{\bf z}_0\}}{{\bf g}^H(\hat{\phi}_{\{{\bf R}_N,{\bf z}_0\}}){\bf C}_N^{-1}{\bf g}(\hat{\phi}_{\{{\bf R}_N,{\bf z}_0\}})}\right|\\ \nonumber
&\leq \max_{\phi\in\mathcal{R}_l}\left|\frac{{\rm Re}^2\{{\bf g}^H(\phi){\bf R}_N^{-1}{\bf z}_0\}}{{\bf g}^H(\phi){\bf R}_N^{-1}{\bf g}(\phi)}\!-\!\frac{1}{1-c}\frac{{\rm Re}^2\{{\bf g}^H(\phi){\bf C}_N^{-1}{\bf z}_0\}}{{\bf g}^H(\phi){\bf C}_N^{-1}{\bf g}(\phi)}\right|\\ \label{ieq:3}
&\stackrel{\rm a.s.}\longrightarrow0, \\\nonumber
&\left|\frac{{\rm Re}^2\{{\bf g}^H(\hat{\phi}){\bf R}_N^{-1}{\bf z}_0\}}{{\bf g}^H(\hat{\phi}){\bf R}_N^{-1}{\bf g}(\hat{\phi})}-\frac{1}{1-c}\frac{{\rm Re}^2\{{\bf g}^H(\hat{\phi}){\bf C}_N^{-1}{\bf z}_0\}}{{\bf g}^H(\hat{\phi}){\bf C}_N^{-1}{\bf g}(\hat{\phi})}\right| \\\nonumber
&\leq \max_{\phi\in\mathcal{R}_l}\left|\frac{{\rm Re}^2\{{\bf g}^H(\phi){\bf R}_N^{-1}{\bf z}_0\}}{{\bf g}^H(\phi){\bf R}_N^{-1}{\bf g}(\phi)}-\frac{1}{1-c}\frac{{\rm Re}^2\{{\bf g}^H(\phi){\bf C}_N^{-1}{\bf z}_0\}}{{\bf g}^H(\phi){\bf C}_N^{-1}{\bf g}(\phi)}\right|\\ \label{ieq:4}
&\stackrel{\rm a.s.}\longrightarrow0,
\end{align}
\normalsize
Using (\ref{ieq:3}) and (\ref{ieq:4}) in (\ref{ieq:1}), it follows that  for all large $N$, almost surely,
\begin{align}\label{ieq:5}
\frac{{\rm Re}^2\{{\bf g}^H(\hat{\phi}){\bf C}_N^{-1}{\bf z}_0\}}{{\bf g}^H(\hat{\phi}){\bf C}_N^{-1}{\bf g}(\hat{\phi})}\leq\frac{{\rm Re}^2\{{\bf g}^H(\hat{\phi}_{\{{\bf R}_N,{\bf z}_0\}}){\bf C}_N^{-1}{\bf z}_0\}}{{\bf g}^H(\hat{\phi}_{\{{\bf R}_N,{\bf z}_0\}}){\bf C}_N^{-1}{\bf g}(\hat{\phi}_{\{{\bf R}_N,{\bf z}_0\}})} \; .
\end{align}
Thus (\ref{ieq:2}) and (\ref{ieq:5}) together ensure that
\small
\begin{align}\nonumber
\left|\frac{{\rm Re}^2\{{\bf g}^H(\hat{\phi}_{\{{\bf R}_N,{\bf z}_0\}}){\bf C}_N^{-1}{\bf z}_0\}}{{\bf g}^H(\hat{\phi}_{\{{\bf R}_N,{\bf z}_0\}}){\bf C}_N^{-1}{\bf g}(\hat{\phi}_{\{{\bf R}_N,{\bf z}_0\}})}-\frac{{\rm Re}^2\{{\bf g}^H(\hat{\phi}){\bf C}_N^{-1}{\bf z}_0\}}{{\bf g}^H(\hat{\phi}){\bf C}_N^{-1}{\bf g}(\hat{\phi})}\right|\stackrel{\rm a.s.}\longrightarrow0 \; .
\end{align}
\normalsize

To complete the proof, we now present the derivations of (\ref{eq:conv_dmn}) and (\ref{eq:conv_nmt}). Since
\begin{align}\nonumber
{\bf R}_N=\frac{1}{K} \sum_{k=1}^K {\bf z}_k{\bf z}_k^H={\bf C}_N^{1/2}\left(\frac{1}{K}\sum_{k=1}^K{\bf q}_k{\bf q}_k^H\right){\bf C}_N^{1/2}, \end{align}
we rewrite $\frac{1}{N}{\bf g}^H(\phi){\bf R}_N^{-1}{\bf g}(\phi)$ as
\begin{align} \nonumber
&\frac{1}{N}{\bf g}^H(\phi){\bf R}_N^{-1}{\bf g}(\phi)\\ \label{eq:rewrite}
&=\frac{1}{N}{\bf g}^H(\phi){\bf C}_N^{-1}{\bf g}(\phi)\tilde{\bf g}^H(\phi)\left(\frac{1}{K}\sum_{k=1}^K{\bf q}_k{\bf q}_k^H\right)^{-1}\tilde{\bf g}(\phi)
\end{align}
where $\tilde{\bf g}(\phi)=\frac{{\bf g}(\phi){\bf C}_N^{-1/2}}{\sqrt{{\bf g}(\phi){\bf C}_N^{-1}{\bf g}(\phi)}}$. \\
 Next, we note that $\tilde{\bf g}^H(\phi)\left(\frac{1}{K}\sum_{k=1}^K{\bf q}_k{\bf q}_k^H\right)^{-1}\tilde{\bf g}(\phi)$ is a rotation invariant scalar, hence we have
 \begin{align}\nonumber
 \tilde{\bf g}^H(\phi)\left(\frac{1}{K}\sum_{k=1}^K{\bf q}_k{\bf q}_k^H\right)^{-1}\tilde{\bf g}(\phi)=\frac{1}{N}{\bf 1}_N^H {\bf \Lambda}^{-1}{\bf 1}_N
 \end{align}
 where ${\bf \Lambda}={\rm diag}(\lambda_1,\ldots,\lambda_N)$ is a diagonal matrix with diagonal entries $\lambda_1,\ldots, \lambda_N$ equal to the eigenvalues of $\frac{1}{K}\sum_{k=1}^K{\bf q}_k{\bf q}_k^H$ \cite{pafka2003noisy}.
Denote the empirical eigenvalue distribution of $\frac{1}{K}\sum_{k=1}^K{\bf q}_k{\bf q}_k^H$ as $f(\lambda)=\frac{1}{N}\sum_{i=1}^N\delta(\lambda-\lambda_i)$
where $\delta(\lambda)$ is the Dirac delta function.
According to the Mar\v{c}enko-Pastur law \cite{marvcenko1967distribution}, as $N, K\rightarrow\infty$, with $c_N=N/K\rightarrow c\in(0,1]$, $f(\lambda)$ converges almost surely to a non-random limiting eigenvalue distribution
\begin{align}\label{eq:conv_lambda}
\rho(\lambda)=\frac{1}{2\pi c}\frac{\sqrt{(\lambda_{+}-\lambda)(\lambda-\lambda_{-})}}{\lambda}, \quad \quad \lambda \in [ \lambda_{-}, \lambda_{+} ]
\end{align}
where $\lambda_{\pm}=(1\pm \sqrt{c})^2$. As a consequence \cite{pafka2003noisy}
\begin{align}\nonumber
\left| \frac{1}{N}{\bf 1}_N^H {\bf \Lambda}^{-1}{\bf 1}_N  -\int\rho(\lambda)/\lambda\, {\rm d}\lambda\right|\stackrel{\rm a.s.}\longrightarrow0 ,  \nonumber
\end{align}
and, equivalently,
\small
\begin{align}\label{eq:conv_dmn_2}
\max_{\phi\in\mathcal{R}_l}\left|\tilde{\bf g}^H(\phi)\left(\frac{1}{K}\sum_{k=1}^K{\bf q}_k{\bf q}_k^H\right)^{-1}\tilde{\bf g}(\phi)\!-\!\int\rho(\lambda)/\lambda\, {\rm d}\lambda\right|\stackrel{\rm a.s.}\longrightarrow0.
\end{align}
\normalsize
Since $\int\rho(\lambda)/\lambda\, {\rm d}\lambda=\frac{1}{1-c}$, combining (\ref{eq:rewrite}) and (\ref{eq:conv_dmn_2}), the convergence (\ref{eq:conv_dmn}) follows.

As for (\ref{eq:conv_nmt}), it should hold under both hypotheses.  Under $H_0$, with ${\bf z}_0={\bf n}_0$, define $\tilde{\bf n}_0={\bf C}_N^{-1/2}{\bf n}_0$. Then we have
\begin{align}\nonumber
&\frac{1}{N}{\rm Re}^2\{{\bf g}^H(\phi){\bf R}_N^{-1}{\bf z}_0\} \\\nonumber
&=\frac{1}{N}{\bf g}^H(\phi){\bf C}_N^{-1}{\bf g}(\phi){\rm Re}^2\left\{\tilde{\bf g}^H(\phi)\left(\frac{1}{K}\sum_{k=1}^K{\bf q}_k{\bf q}_k^H\right)^{-1}\tilde{\bf n}_0\right\}.
\end{align}
Again, since $\frac{1}{K}\sum_{k=1}^K{\bf q}_k{\bf q}_k^H$ is rotation invariant, we have
\begin{align}\nonumber
\max_{\phi\in\mathcal{R}_l}&\left|\frac{1}{N}{\rm Re}^2\{{\bf g}^H(\phi){\bf R}_N^{-1}{\bf z}_0\}\right.\\ \nonumber
&~\left.-\frac{1}{2N}{\bf g}^H(\phi){\bf C}_N^{-1}{\bf g}(\phi)\left(\int\rho(\lambda)/\lambda\, {\rm d}\lambda\right)^2\right|\stackrel{\rm a.s.}\longrightarrow0.
\end{align}
Thus,
\begin{align} \nonumber
\max_{\phi\in\mathcal{R}_l}&\left|\frac{1}{N}{\rm Re}^2\{{\bf g}^H(\phi){\bf R}_N^{-1}{\bf z}_0\}\right.\\\label{eq:conv_nmt_1}
&~\left.-\frac{1}{(1-c)^2}\frac{1}{2N}{\bf g}^H(\phi){\bf C}_N^{-1}{\bf g}(\phi)\right|\stackrel{\rm a.s.}\longrightarrow0.
\end{align}
Since also
\begin{align}\label{eq:conv_nmt_2}
\max_{\phi\in\mathcal{R}_l}\!\left|\frac{1}{N}{\rm Re}^2\{{\bf g}^H(\phi){\bf C}_N^{-1}{\bf z}_0\}\!-\!\frac{1}{2N}{\bf g}^H(\phi){\bf C}_N^{-1}{\bf g}(\phi)\right|\!\!\stackrel{\rm a.s.}\longrightarrow\!0,
\end{align}
 combining (\ref{eq:conv_nmt_1}) and (\ref{eq:conv_nmt_2}), we obtain the convergence (\ref{eq:conv_nmt}) when ${\bf z}_0={\bf n}_0$.

 Under $H_1$, with ${\bf z}_0=s{\bf g}(\phi)+{\bf n}_0$, we have
 \begin{align}\nonumber
 {\bf g}^H(\phi){\bf R}_N^{-1}{\bf z}_0=s{\bf g}^H(\phi){\bf R}_N^{-1}{\bf g}(\phi)+{\bf g}^H(\phi){\bf R}_N^{-1}{\bf n}_0.
 \end{align}
 Relating to (\ref{eq:conv_dmn}) and (\ref{eq:conv_nmt_1}), we obtain
 \begin{align}\nonumber
 \max_{\phi\in\mathcal{R}_l}&\left|\frac{1}{N}{\rm Re}^2\{{\bf g}^H(\phi){\bf R}_N^{-1}{\bf z}_0\}\right.\\\label{eq:conv_nmt1_H1}
 &~\left.-\frac{2s^2+1}{2N(1-c)^2}{\bf g}^H(\phi){\bf C}_N^{-1}{\bf g}(\phi)\right|\stackrel{\rm a.s.}\longrightarrow0.
 \end{align}
 \small
 \begin{align}\label{eq:conv_nmt1_H1}
 \max_{\phi\in\mathcal{R}_l}\left|\frac{1}{N}{\rm Re}^2\{{\bf g}^H(\phi){\bf R}_N^{-1}{\bf z}_0\}\!-\!\frac{2s^2+1}{2N(1-c)^2}{\bf g}^H(\phi){\bf C}_N^{-1}{\bf g}(\phi)\right|\!\!\stackrel{\rm a.s.}\longrightarrow\!0.
 \end{align}
 \normalsize
 Similarly,
 \small
 \begin{align}\nonumber
  \max_{\phi\in\mathcal{R}_l}&\left|\frac{1}{N}{\rm Re}^2\{{\bf g}^H(\phi){\bf C}_N^{-1}{\bf z}_0\}\!-\!\frac{2s^2+1}{2N}{\bf g}^H(\phi){\bf C}_N^{-1}{\bf g}(\phi)\right|\!\stackrel{\rm a.s.}\longrightarrow0,
 \end{align}
 \normalsize
 which together with (\ref{eq:conv_nmt1_H1}) further yields (\ref{eq:conv_nmt}).

\end{appendices}

\section*{acknowledgement}
The authors thank Mohamed S. Ghidaoui of HKUST's Department of Civil and Environmental Engineering for numerous helpful discussions throughout the course of this work, particularly relating to details of the physical water pipeline model and associated practicalities.

\bibliographystyle{IEEEtran}
\bibliography{cited}

\begin{thebibliography}{10}
\providecommand{\url}[1]{#1}
\csname url@samestyle\endcsname
\providecommand{\newblock}{\relax}
\providecommand{\bibinfo}[2]{#2}
\providecommand{\BIBentrySTDinterwordspacing}{\spaceskip=0pt\relax}
\providecommand{\BIBentryALTinterwordstretchfactor}{4}
\providecommand{\BIBentryALTinterwordspacing}{\spaceskip=\fontdimen2\font plus
\BIBentryALTinterwordstretchfactor\fontdimen3\font minus
  \fontdimen4\font\relax}
\providecommand{\BIBforeignlanguage}[2]{{%
\expandafter\ifx\csname l@#1\endcsname\relax
\typeout{** WARNING: IEEEtran.bst: No hyphenation pattern has been}%
\typeout{** loaded for the language `#1'. Using the pattern for}%
\typeout{** the default language instead.}%
\else
\language=\csname l@#1\endcsname
\fi
#2}}
\providecommand{\BIBdecl}{\relax}
\BIBdecl

\bibitem{ghazali2012comparative}
M.~Ghazali, S.~Beck, J.~Shucksmith, J.~Boxall, and W.~Staszewski, ``Comparative
  study of instantaneous frequency based methods for leak detection in pipeline
  networks,'' \emph{Mech. Syst. Signal Process.}, vol.~29, pp. 187--200, May
  2012.

\bibitem{vitkovsky2000leak}
J.~P. V{\'\i}tkovsk{\`y}, A.~R. Simpson, and M.~F. Lambert, ``Leak detection
  and calibration using transients and genetic algorithms,'' \emph{J. Water
  Resource Plan. Manag.}, vol. 126, no.~4, pp. 262--265, 2000.

\bibitem{wang2018pipeline}
X.~Wang and M.~S. Ghidaoui, ``Pipeline leak detection using the matched-field
  processing method,'' \emph{J. Hydraulic Eng.}, vol. 144, no.~6, p. 04018030,
  Jun. 2018.

\bibitem{wang2018identification}
------, ``Identification of multiple leaks in pipeline: Linearized model,
  maximum likelihood, and super-resolution localization,'' \emph{Mech. Syst.
  Signal Process.}, vol. 107, pp. 529--548, Jul. 2018.

\bibitem{wang2018experiment}
X.~Wang, J.~Lin, A.~Keramat, M.~S. Ghidaoui, S.~Meniconi, and B.~Brunone,
  ``Matched-field processing for leak detection in a viscoelasticity pipe: {A}n
  experimental study,'' \emph{Mech. Syst. Signal Process.}, vol. 124, pp.
  459--478, 2019.

\bibitem{kelly1986adaptive}
E.~J. Kelly, ``An adaptive detection algorithm,'' \emph{IEEE Tran. Aerosp.
  Electron. Syst.}, no.~2, pp. 115--127, 1986.

\bibitem{ledoit2004well}
O.~Ledoit and M.~Wolf, ``A well-conditioned estimator for large-dimensional
  covariance matrices,'' \emph{J. Multivariate Anal.}, vol.~88, no.~2, pp.
  365--411, Feb. 2004.

\bibitem{rubio2012performance}
F.~Rubio, X.~Mestre, and D.~P. Palomar, ``Performance analysis and optimal
  selection of large minimum variance portfolios under estimation risk,''
  \emph{IEEE J. Sel. Topics Signal Process.}, vol.~6, no.~4, pp. 337--350, Aug.
  2012.

\bibitem{abramovich1981controlled}
Y.~I. Abramovich, ``A controlled method for adaptive optimization of filters
  using the criterion of maximum signal-to-noise ratio,'' \emph{Radio Eng.
  Elect. Phys}, vol.~26, no.~3, pp. 87--95, 1981.

\bibitem{abramovich2007modified}
Y.~I. Abramovich, N.~K. Spencer, and A.~Y. Gorokhov, ``Modified {GLRT} and
  {AMF} framework for adaptive detectors,'' \emph{IEEE Trans. Aerosp. Electron.
  Syst.}, vol.~43, no.~3, Jul. 2007.

\bibitem{mestre2006finite}
X.~Mestre and M.~{\'A}. Lagunas, ``Finite sample size effect on minimum
  variance beamformers: Optimum diagonal loading factor for large arrays,''
  \emph{IEEE Trans. Signal Process.}, vol.~54, no.~1, pp. 69--82, 2006.

\bibitem{carlson1988covariance}
B.~D. Carlson, ``Covariance matrix estimation errors and diagonal loading in
  adaptive arrays,'' \emph{IEEE Tran. Aerosp. Electron. Syst.}, vol.~24, no.~4,
  pp. 397--401, 1988.

\bibitem{couillet2014large}
R.~Couillet and M.~R. McKay, ``Large dimensional analysis and optimization of
  robust shrinkage covariance matrix estimators,'' \emph{J. Mult. Anal.}, vol.
  131, pp. 99--120, 2014.

\bibitem{yang2015robust}
L.~Yang, R.~Couillet, and M.~R. McKay, ``A robust statistics approach to
  minimum variance portfolio optimization,'' \emph{IEEE Trans. Signal
  Process.}, vol.~63, no.~24, pp. 6684--6697, Aug. 2015.

\bibitem{auguin2018large}
N.~Auguin, D.~Morales-Jimenez, M.~R. McKay, and R.~Couillet,
  ``Large-dimensional behavior of regularized {M}aronna's {M}-estimators of
  covariance matrices,'' \emph{IEEE Trans. Signal Process.}, vol.~66, no.~13,
  pp. 3529--3542, 2018.

\bibitem{ma2003efficient}
N.~Ma and J.~T. Goh, ``Efficient method to determine diagonal loading value,''
  in \emph{Proc. IEEE Int. Conf. Acoust., Speech, Signal Process.}, vol.~5,
  2003, pp. 341--344.

\bibitem{kammoun2018optimal}
A.~Kammoun, R.~Couillet, F.~Pascal, and M.~Alouini, ``Optimal design of the
  adaptive normalized matched filter detector using regularized {T}yler
  estimators,'' \emph{IEEE Trans. Aerosp. Electron. Syst.}, vol.~54, no.~2, pp.
  755--769, Apr. 2018.

\bibitem{couillet2016second}
R.~Couillet, A.~Kammoun, and F.~Pascal, ``Second order statistics of robust
  estimators of scatter: {A}pplication to {GLRT} detection for elliptical
  signals,'' \emph{J. Multivariate Anal.}, vol. 143, pp. 249--274, Jan. 2016.

\bibitem{dubey2019measurement}
A.~Dubey, Z.~Li, P.~Lee, and R.~Murch, ``Measurement and characterization of
  acoustic noise in water pipeline channels,'' \emph{IEEE Access}, vol.~7, pp.
  56\,890--56\,903, 2019.

\bibitem{lehmann2006testing}
E.~L. Lehmann and J.~P. Romano, \emph{Testing Statistical Hypotheses}.\hskip
  1em plus 0.5em minus 0.4em\relax Springer Science \& Business Media, 2006.

\bibitem{goodman1963statistical}
N.~R. Goodman, ``Statistical analysis based on a certain multivariate complex
  {G}aussian distribution (an introduction),'' \emph{Ann. Statist.}, vol.~34,
  no.~1, pp. 152--177, 1963.

\bibitem{raghavan1995cfar}
R.~Raghavan, H.~Qiu, and D.~McLaughlin, ``{CFAR} detection in clutter with
  unknown correlation properties,'' \emph{IEEE Trans. Aerosp. Electron. Syst.},
  vol.~31, no.~2, pp. 647--657, 1995.

\bibitem{reed1974rapid}
I.~S. Reed, J.~D. Mallett, and L.~E. Brennan, ``Rapid convergence rate in
  adaptive arrays,'' \emph{IEEE Tran. Aerosp. Electron. Syst.}, no.~6, pp.
  853--863, 1974.

\bibitem{boroson1980sample}
D.~M. Boroson, ``Sample size considerations for adaptive arrays,'' \emph{IEEE
  Trans. Aerosp. Electron. Syst.}, no.~4, pp. 446--451, 1980.

\bibitem{mestre2008improved}
X.~Mestre, ``Improved estimation of eigenvalues and eigenvectors of covariance
  matrices using their sample estimates,'' \emph{IEEE Trans. Inf. Theory},
  vol.~54, no.~11, pp. 5113--5129, Nov. 2008.

\bibitem{marvcenko1967distribution}
V.~A. Mar{\v{c}}enko and L.~A. Pastur, ``Distribution of eigenvalues for some
  sets of random matrices,'' \emph{Math. USSR-Sbornik}, vol.~1, no.~4, p. 457,
  1967.

\bibitem{mestre2008asymptotic}
X.~Mestre, ``On the asymptotic behavior of the sample estimates of eigenvalues
  and eigenvectors of covariance matrices,'' \emph{IEEE Trans. Signal
  Process.}, vol.~56, no.~11, pp. 5353--5368, 2008.

\bibitem{silverstein1995strong}
J.~W. Silverstein, ``Strong convergence of the empirical distribution of
  eigenvalues of large dimensional random matrices,'' \emph{J. Multivar.
  Anal.}, vol.~55, no.~2, pp. 331--339, 1995.

\bibitem{silverstein1995empirical}
J.~W. Silverstein and Z.~Bai, ``On the empirical distribution of eigenvalues of
  a class of large dimensional random matrices,'' \emph{J. Multivariate Anal.},
  vol.~54, no.~2, pp. 175--192, 1995.

\bibitem{gini1997sub}
F.~Gini, ``Sub-optimum coherent radar detection in a mixture of {K}-distributed
  and {G}aussian clutter,'' \emph{IEE Proc. Radar, Sonar, Navig.}, vol. 144,
  no.~1, pp. 39--48, 1997.

\bibitem{wylie1993fluid}
E.~B. Wylie, V.~L. Streeter, and L.~Suo, \emph{Fluid {T}ransients in
  {S}ystems}.\hskip 1em plus 0.5em minus 0.4em\relax Prentice Hall Englewood
  Cliffs, NJ, 1993, vol.~1.

\bibitem{chaudhry1979applied}
M.~H. Chaudhry, \emph{Applied {H}ydraulic {T}ransients}, {T}hird~ed., 2014.

\bibitem{pozar2009microwave}
D.~M. Pozar, \emph{Microwave Engineering}.\hskip 1em plus 0.5em minus
  0.4em\relax John Wiley \& Sons, 2009.

\bibitem{shah1998performance}
A.~Shah and A.~M. Haimovich, ``Performance analysis of optimum combining in
  wireless communications with {R}ayleigh fading and cochannel interference,''
  \emph{IEEE Trans. Commun.}, vol.~46, no.~4, pp. 473--479, 1998.

\bibitem{gut2013probability}
A.~Gut, \emph{Probability: {A} {G}raduate {C}ourse}.\hskip 1em plus 0.5em minus
  0.4em\relax Springer Science \& Business Media, 2013, vol.~75.

\bibitem{kallenberg1997foundations}
O.~Kallenberg and O.~Kallenberg, \emph{Foundations of modern
  probability}.\hskip 1em plus 0.5em minus 0.4em\relax Springer, 1997, vol.~2.

\bibitem{pafka2003noisy}
S.~Pafka and I.~Kondor, ``Noisy covariance matrices and portfolio optimization
  {II},'' \emph{Phys. A}, vol. 319, pp. 487--494, 2003.

\end{thebibliography}

\end{document}